\documentclass[a4paper,fleqn,usenatbib]{mnras}

\usepackage{newtxtext,newtxmath}

\usepackage[T1]{fontenc}
\usepackage{ae,aecompl}

\usepackage{graphicx}	
\usepackage{amsmath}	
\usepackage{amssymb}	
\usepackage{hyperref}
\usepackage{caption} 

\usepackage{soul} 

\newcommand{\todo}[1]{{#1}}
\newcommand{\inprep}[1]{\textcolor{blue}{#1}}

\title[Kinematics of strong and weak bars]{Galaxy Zoo: Kinematics of strongly and weakly barred galaxies}

\author[Tobias G\'eron]{
Tobias G\'eron$^{1}$\thanks{E-mail: tobias.geron@physics.ox.ac.uk (TG)},
Rebecca J. Smethurst$^{1}$,
Chris Lintott$^{1}$,
Sandor Kruk$^{2}$,
Karen L. Masters$^{3}$,
\newauthor
Brooke Simmons$^{4}$,
Kameswara Bharadwaj Mantha$^{5, 6}$,
Mike Walmsley$^{7}$,
\newauthor
L. Garma-Oehmichen$^{8}$,
Niv Drory$^{9}$,
Richard R. Lane$^{10}$
\\
$^{1}$ Oxford Astrophysics, Department of Physics, University of Oxford, Denys Wilkinson Building, Keble Road, Oxford, OX1 3RH, UK\\
$^{2}$ Max-Planck-Institut f\"ur extraterrestrische Physik (MPE), Giessenbachstrasse 1, D-85748 Garching bei M\"unchen, Germany\\
$^{3}$ Haverford College, Department of Physics and Astronomy, 370 Lancaster Avenue, Haverford, Pennsylvania 19041, USA\\
$^{4}$ Department of Physics, Lancaster University, Lancaster, LA1 4YB, UK\\
$^{5}$ Minnesota Institute for Astrophysics, University of Minnesota, 116 Church St SE, Minneapolis, MN 55455, USA\\
$^{6}$ Department of Physics and Astronomy, University of Minnesota, 116 Church St SE, Minneapolis, MN 55455, USA\\
$^{7}$ Jodrell Bank Centre for Astrophysics, Department of Physics \& Astronomy, University of Manchester, Oxford Road, Manchester M13 9PL, UK\\
$^{8}$ Instituto de Astronom\'ia, Universidad Nacional Aut\'onoma de M\'exico, Apartado Postal 70-264, CDMX, 04510, M\'exico\\
$^{9}$ McDonald Observatory, The University of Texas at Austin, 1 University Station, Austin, TX 78712, USA\\
$^{10}$ Centro de Investigaci\'on en Astronom\'ia, Universidad Bernardo O'Higgins, Avenida Viel 1497, Santiago, Chile\\
}

\date{Accepted XXX. Received YYY; in original form ZZZ}

\pubyear{2022}

\begin{document}
\label{firstpage}
\pagerange{\pageref{firstpage}--\pageref{lastpage}}
\maketitle

\begin{abstract}

    We study the bar pattern speeds and corotation radii of \todo{225} barred galaxies, using IFU data from MaNGA and the Tremaine-Weinberg method. Our sample, which is divided between strongly and weakly barred galaxies identified via Galaxy Zoo, is the largest that this method has been applied to. We find lower pattern speeds for strongly barred galaxies than for weakly barred galaxies. As simulations show that the pattern speed decreases as the bar exchanges angular momentum with its host, these results suggest that strong bars are more evolved than weak bars. Interestingly, the corotation radius is not different between weakly and strongly barred galaxies, despite being proportional to bar length. We also find that the corotation radius is significantly different between quenching and star forming galaxies. Additionally, we find that strongly barred galaxies have significantly lower values for $\mathcal{R}$, the ratio between the corotation radius and the bar radius, than weakly barred galaxies, despite a big overlap in both distributions. This ratio classifies bars into ultrafast bars ($\mathcal{R} < $ 1.0; \todo{11}\% of our sample), fast bars (1.0 $< \mathcal{R} <$ 1.4; \todo{27}\%) and slow bars ($\mathcal{R} >$ 1.4; \todo{62}\%). Simulations show that $\mathcal{R}$ is correlated with the bar formation mechanism, so our results suggest that strong bars are more likely to be formed by different mechanisms than weak bars. Finally, we find a lower fraction of ultrafast bars than most other studies, which decreases the recently claimed tension with $\Lambda$CDM. However, the median value of $\mathcal{R}$ is still lower than what is predicted by simulations.
\end{abstract}







\section{Introduction}

Bars are a relatively common structure in galaxies, with about 30\%-60\% of nearby galaxies hosting a bar, depending on the redshift and wavelength range of the study \citep{marinova_2007, menendez_delmestre_2007, barazza_2008, sheth_2008, nair_2010b, masters_2011}. Bars can drive angular momentum outwards and funnel gas to the centre of the galaxy \citep{athanassoula_1992, davoust_2004, rodriguez_fernandez_2008, athanassoula_2013, villa-vargas_2010, fragkoudi_2016, vera_2016, spinoso_2017, george_2019,seo_2019}. This will result in significant ``secular'' evolution of the host, caused directly by its bar \citep{kormendy_2004,cheung_2013,sellwood_2014,diaz_garcia_2016b,kruk_2018}. Moreover, multiple studies have found that bars appear more often in massive, red and gas-poor galaxies \citep{masters_2012,cervantessodi_2017,vera_2016,fraser_mckelvie_2020b}. For example, \citet{kruk_2018} found that bars are redder than disks and that the disks of barred galaxies are redder than the disks in unbarred galaxies. These results suggest that bars might be linked to the quenching of their host. This can be either the result of triggering a starburst in the centre of the galaxy, after extensive inflows of gas \citep{alonso_herrero_2001, sheth_2005, jogee_2005, hunt_2008, carles_2016}, or by making the gas too dynamically hot for star formation \citep{zurita_2004, haywood_2016,khoperskov_2018,athanassoula_1992,reynaud_1998,sheth_2000}. In any case, a bar is a very common and important structure in a galaxy, so understanding bars is fundamental to understanding galaxy evolution.

Bars are historically classified into weak or strong. Since \citet{devaucouleurs_1959,devaucouleurs_1963}, three subclasses are recognised: unbarred (SA), strongly barred (SB) and weakly barred (SAB). Weakly barred galaxies were thought to be an intermediate class between unbarred and strongly barred, having lengths and contrast in between SA and SB bars. Bars classified as weak were usually small and faint, whereas bars classified as strong were long and obvious \citet{devaucouleurs_1959, devaucouleurs_1963}. Morphological arguments are still used to determine bar type. \citet{nair_2010} produced a catalogue of detailed visual morphological classifications and they distinguished between weak and strong by looking at whether the bar dominated the light distribution. The bar strength can also be estimated using the maximum ellipticity and boxiness of the isophotes \citep{athanassoula_1992b,laurikainen_2002,erwin_2004,gadotti_2011}. One can also look at the surface brightness profiles of bars. Previous work has shown that stronger bars have flat profiles, while weaker bars have exponential profiles \citep{elmegreen_1985, elmegreen_1996, kim_2015, kruk_2018}. Clearly, there are many ways to characterise bars and their strength. However, the community has yet to reach a consensus on how to best define weak and strong bars and on which detection method is superior. 

This problem was addressed more recently by the Galaxy Zoo (GZ) team, who combined the efforts of citizen scientists and machine learning to provide morphological classifications of galaxies \citep{lintott_2008,walmsley_2022}. These morphological classifications included a distinction between weak and strong bars based on visual morphology. Volunteers are shown examples of weak and strong bars prior to classification. The strong bars are typically large and obvious structures, whereas the weak bars can be smaller and fainter. \citet{geron_2021} used the morphological classifications from GZ based on images from the Dark Energy Camera Legacy Survey (DECaLS, \citealp{dey_2019}) to study weak and strong bars, and found that around 28\% of all disk galaxies have a weak bar, while 16\% had a strong bar. They also found that, when correcting for bar length, any difference they observed between weak and strong bars disappeared. Thus, they suggested that weak and strong bars are not fundamentally different physical phenomena. Instead, they proposed the existence of a continuum of bar types, which varies from `weakest' to `strongest'. Most research on bars has traditionally been focussed on stronger bars, as they are more obvious and clearer structures. However, weaker bars are still very common structures in galaxies and need to be included in more studies to obtain a more complete picture. 

The bar pattern speed ($\Omega_{\rm bar}$), or the rotational frequency of the bar, is one of the most important dynamical parameters that describe a bar. It is intrinsically linked to the evolution of the bar and its host. It is typically found that, as the bar exchanges angular momentum with its host, the bar grows and the pattern speed decreases \citep{debattista_2000,athanassoula_2003, martinez_valpuesta_2006,okamoto_2015}.

If the bar pattern speed and galaxy kinematics are known, one can calculate the corotation radius (R$_{\rm CR}$), which is the radius at which the angular speed of the stars in the disc is equal to the pattern speed of the bar. Additionally, one can also calculate the dimensionless corotation radius-to-bar radius ratio, $\mathcal{R}$. A large value for $\mathcal{R}$ implies that the point of corotation is far outside the bar region. This ratio is typically used to separate bars into `fast' (1.0 $<$ $\mathcal{R}$ $<$ 1.4) and `slow' ($\mathcal{R}$ $>$ 1.4) bars \citep{debattista_2000,rautiainen_2008,aguerri_2015}. There is a known correlation between the formation of the bar and $\mathcal{R}$. Bars that are triggered by tidal interactions tend to be in the slow regime for a longer time and have higher values for $\mathcal{R}$ than bars formed by global bar instabilities \citep{sellwood_1981,miwa_1998,martinez_valpuesta_2016,martinez_valpuesta_2017}.

There is also a known tension between simulations and observations on the distribution of the ratio $\mathcal{R}$. Cosmological simulations predict that bars slow down significantly due to dynamical friction with their dark matter halo, which results in high values for $\mathcal{R}$. However, observations typically find lower values of $\mathcal{R}$, which has been highlighted as a challenge for the $\Lambda$CDM cosmology used in these cosmological simulations \citep{algorry_2017,peschken_2019,roshan_2021b}. 

It is suggested that bars cannot extend beyond their corotation radius \citep{contopoulos_1980,contopoulos_1981,athanassoula_1992}. This implies that bars with $\mathcal{R}$ $<$ 1 should not exist. However, these so-called `ultrafast' bars have been repeatedly observed \citep{buta_2009, aguerri_2015, cuomo_2019,guo_2019,garma_oehmichen_2020,krishnarao_2022}. This discrepancy between observation and theory remains an open question, although some suggest that the cause for this problem is rooted in incorrect estimates of the bar radius \citep{cuomo_2021, roshan_2021a}.

It is becoming clear that the pattern speed and the parameters derived from it (such as corotation radius and $\mathcal{R}$) are important. However, it is also quite challenging to correctly estimate the bar pattern speed. Nevertheless, various methods exist to measure this dynamical parameter. For example, one can match the observed surface gas distribution or gas velocity field with simulations where $\Omega_{\rm bar}$ is a free parameter \citep{sanders_1980,hunter_1988,lindblad_1996,weiner_2001,rautiainen_2008,treuthardt_2008}. Alternatively, one can subtract a rotation model from the gas velocity field and look at the morphology of the residuals to estimate the pattern speed \citep{sempere_1995,font_2011,font_2017}. Other morphological features are helpful to determine the bar pattern speed, such as rings \citep{buta_1986, rautiainen_2000,munoz_tunon_2004,perez_2012}, the shape and offset of dust lanes \citep{athanassoula_1992, sanchez_menguiano_2015} and the morphology of spiral arms \citep{puerari_1997, aguerri_1998, buta_2009,sierra_2015}. 

However, all these methods require some sort of modelling. The only reliable direct and model-independent method to determine the bar pattern speed is the Tremaine-Weinberg (TW) method \citep{tremaine_1984}. It has been used extensively in the past to study bar pattern speeds \citep{aguerri_2015,cuomo_2019,guo_2019, garma_oehmichen_2020}. The TW method uses surface brightness and line-of-sight (LOS) velocity data to estimate the pattern speed. 

In this paper, we use the TW method on integral-field spectroscopy data from the Mapping Nearby Galaxies at Apache Point Observatory (MaNGA) survey \citep{bundy_2015} to estimate bar pattern speeds, corotation radii and the dimensionless ratio $\mathcal{R}$ for a sample of \todo{225} galaxies. This is the largest sample to date measured with the TW method and includes both weakly and strongly barred galaxies, identified using GZ.

The structure of the paper is as follows: in Section \ref{sec:tw_method}, we explain the Tremaine-Weinberg method in detail. The data and sample selection is explained in Section \ref{sec:data}. Section \ref{sec:results} shows our results, which are discussed in Section \ref{sec:discussion}. Finally, our conclusions are summarised in Section \ref{sec:conclusion}. Where necessary, we assumed a standard flat cosmological model with H$_{0}$ = 70 km s$^{-1}$ Mpc$^{-1}$, $\Omega_{\rm m}$ = 0.3 and $\Omega_{\rm \Lambda}$ = 0.7.


\section{The Tremaine-Weinberg Method}
\label{sec:tw_method}

\subsection{Theory}
\label{sec:tw_theory}

\begin{figure}
	\includegraphics[width=\columnwidth]{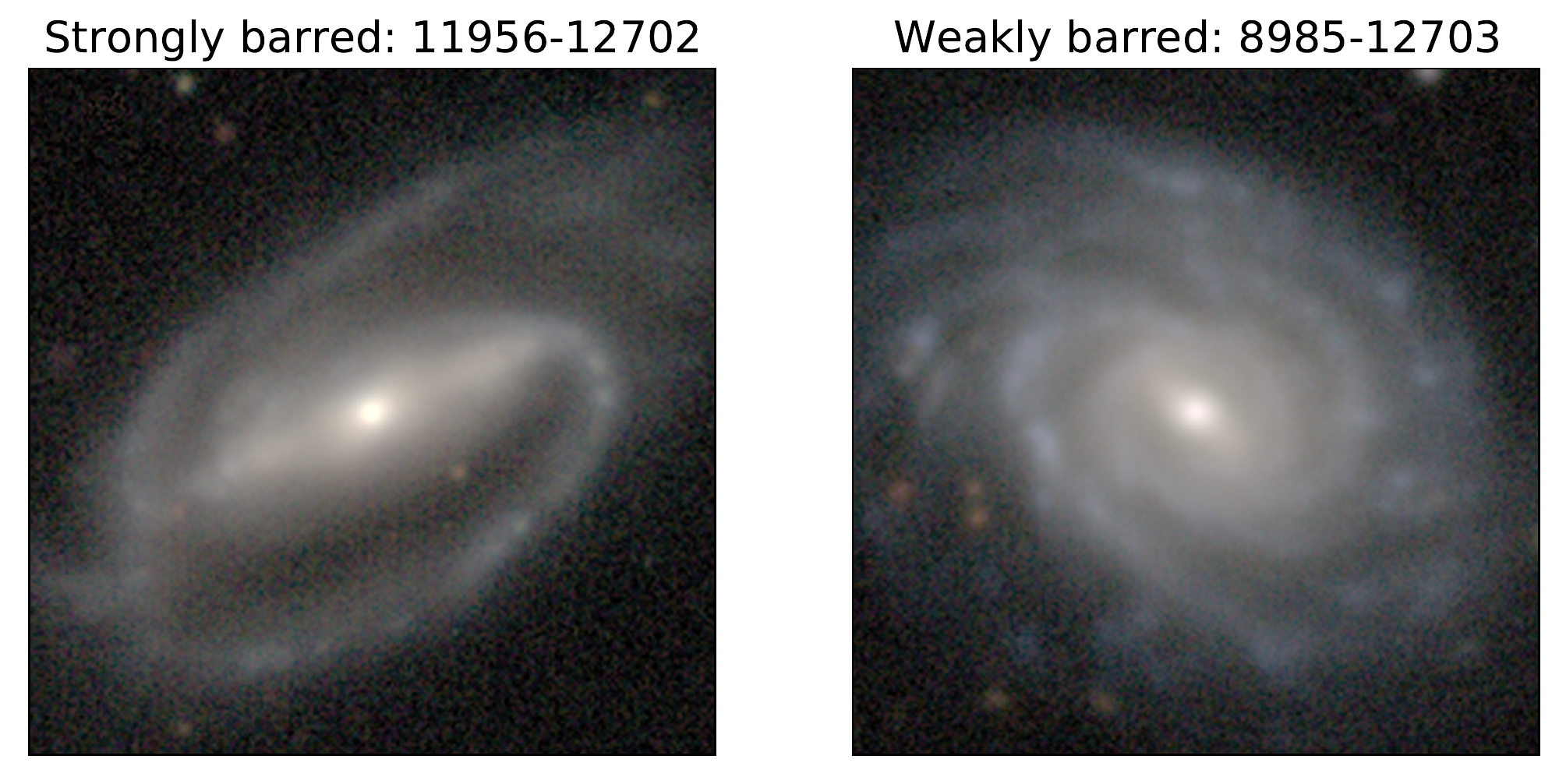}
    \caption{
        DECaLS postage stamps (64x64 arcsec) of a strongly (left) and weakly barred galaxy (right), on which we will apply the Tremaine-Weinberg method.}
    \label{fig:example_decals}
\end{figure}

The Tremaine-Weinberg (TW) method is a model-independent method to determine the pattern speed of a galaxy \citep{tremaine_1984}. The main assumptions of the TW method are that there is a well-defined pattern speed and that the tracer used (i.e. stars or gas) satisfies the continuity equation. To illustrate the different steps of the TW method, we will apply the TW method to one strongly barred galaxy and one weakly barred galaxy, shown in Figure \ref{fig:example_decals}. 

Take a Cartesian coordinate system $\left(X,Y\right)$ in the sky plane with the origin in the centre of the galaxy and the $X$-axis aligned with the line of nodes (LON), which is defined as the intersection of the sky plane and the disc plane, so it is effectively the major axis of the galaxy. Then, the Tremaine-Weinberg method can be formulated as:

\begin{equation}
    \Omega_{\rm b} \sin{\left( i \right)} = \frac{\int_{\rm -\infty}^{\rm +\infty} h(Y) \: \int_{\rm -\infty}^{\rm +\infty} \Sigma(X,Y) \: V_{\rm LOS} (X,Y) \: \textrm{d}X \textrm{d}Y}{\int_{\rm -\infty}^{\rm +\infty} h(Y) \: \int_{\rm -\infty}^{\rm +\infty} X \: \Sigma(X,Y) \: \textrm{d}X \textrm{d}Y} \;,
    \label{eq:tw_full}
\end{equation}

where $\Omega_{\rm b}$ is the bar pattern speed, $i$ is the inclination of the galaxy, $V_{\rm LOS}$ is the line of sight velocity, $\Sigma$ is the surface brightness of the galaxy and $h(Y)$ is a weight function. A delta function like $h(Y) = \delta \left( Y - Y_{0} \right)$ is typically used here, so that the integration happens in pseudo-slits across the IFU parallel to the LON. Multiple integrations are usually done with different offset distances $Y_{0}$ to ensure reliable measurement of the pattern speed \citep{tremaine_1984}. In this case, Equation \ref{eq:tw_full} can be simplified to: 

\begin{equation}
    \Omega_{\rm b} \sin{\left( i \right)} = \frac{\left<V\right>}{\left<X\right>} \;,
\end{equation}

where $\left<X\right>$ is called the photometric integral and $\left<V\right>$ the kinematic integral. They are defined as: 

\begin{equation}
    \left<X\right> = \frac{\int_{\rm -\infty}^{\rm +\infty} X \Sigma \text{d}\Sigma}{\int_{\rm -\infty}^{\rm +\infty} \Sigma \text{d}\Sigma} \, ;  \; \left<V\right> = \frac{\int_{\rm -\infty}^{\rm +\infty} V_{\rm LOS} \Sigma \text{d}\Sigma}{\int_{\rm -\infty}^{\rm +\infty} \Sigma \text{d}\Sigma} \; .
    \label{eq:tw_X_V_eqs}
\end{equation}

\begin{figure}
	\includegraphics[width=\columnwidth]{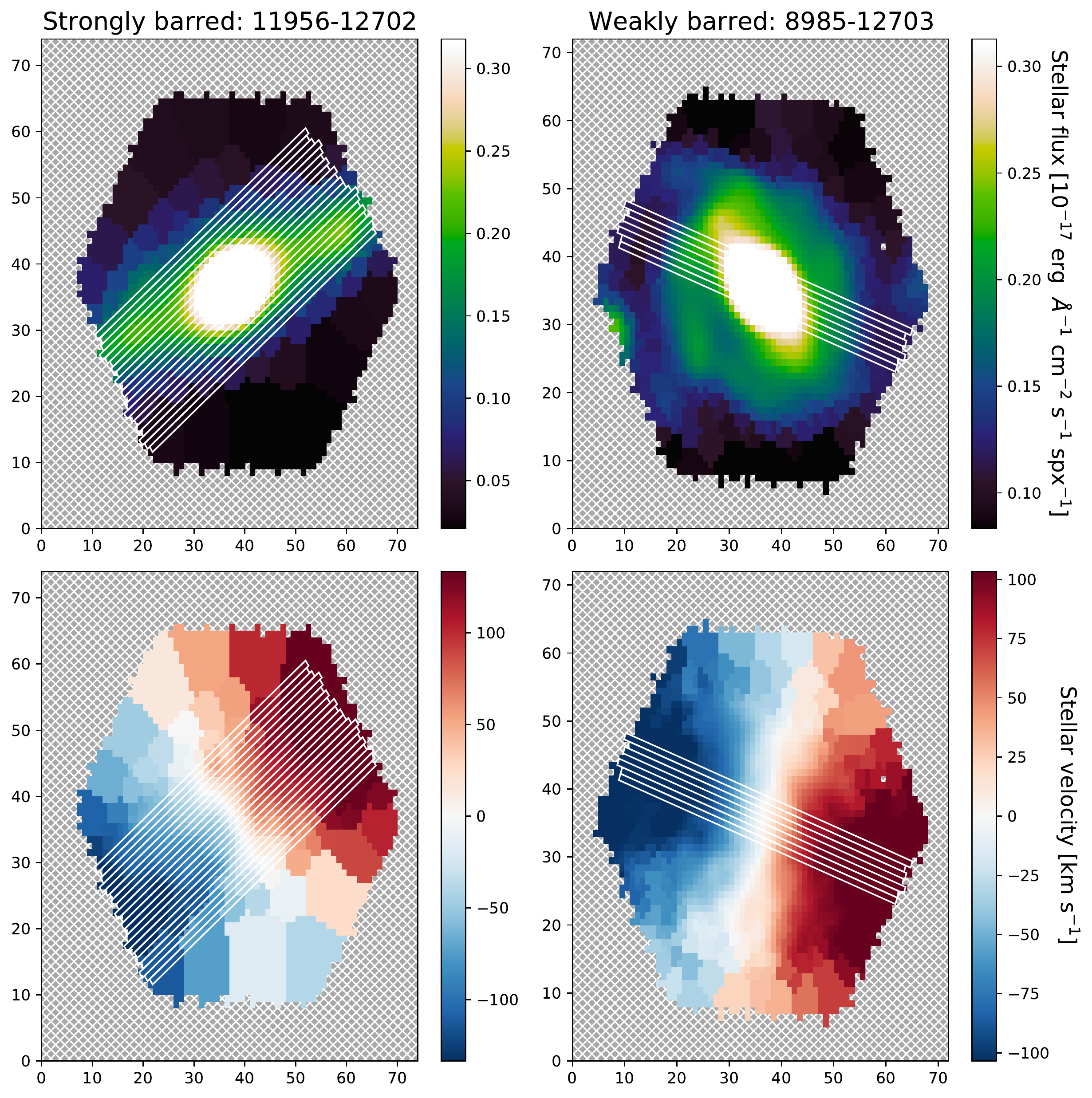}
    \caption{The stellar flux (top row) and stellar velocity (bottom row) for our strongly barred (left column) and weakly barred (right column) example galaxies. The different pseudo-slits, over which the kinematic and photometric integrals are calculated, are visualised on top of the maps in white outlines.}
    \label{fig:example_mangamaps}
\end{figure}

$\left<X\right>$ is effectively the luminosity-weighted mean position and $\left<V\right>$ is the luminosity-weighted mean line of sight velocity. These photometric and kinematic integrals are calculated for the multiple different pseudo-slits across the IFU. These pseudo-slits are visualised on top of the MaNGA stellar flux and velocity maps in Figure \ref{fig:example_mangamaps}. Every pseudo-slit is carefully placed next to each other so that they do not overlap. Every slit has a width of \todo{0.5} arcsec, which is the same width that was used in \citet{guo_2019}. Using different slit widths does not have a significant impact on the final measurement \citep{guo_2019, zou_2019}. To make optimal use of the data, we make the pseudo-slits as long as the data allows. However, each slit should be centred on the disc minor axis. This implies that a target can have multiple slits with sightly different slit lengths. However, this variation is minimal and this approach is similar to what is done in \citet{garma_oehmichen_2020,garma_oehmichen_2022}. We place as many slits as we can fit within the bar, but impose a minimum of \todo{three} slits. The maximum amount of slits placed on one galaxy was \todo{48}, and the median is \todo{10}.

The limits of the integration technically go from $-\infty$ to $+\infty$, as shown in Equation \ref{eq:tw_X_V_eqs}. However, this which is not possible with real data. To make sure that the pseudo-slits are long enough, we test the convergence of every pseudo-slit, as suggested by \citet{zou_2019} and \inprep{Zou et al.} (\inprep{in prep.}). This is done by increasing the length of each pseudo-slit by 1 pixel until its maximum length is reached. A slit has converged if the median of the absolute value of the change in $\Omega_{\rm b} \sin{\left( i \right)}$ in the last \todo{5} slit lengths tested was less than \todo{1} km s$^{-1}$ arcsec$^{-1}$. Any pseudo-slit that did not meet this threshold was discarded.

\begin{figure}
	\includegraphics[width=\columnwidth]{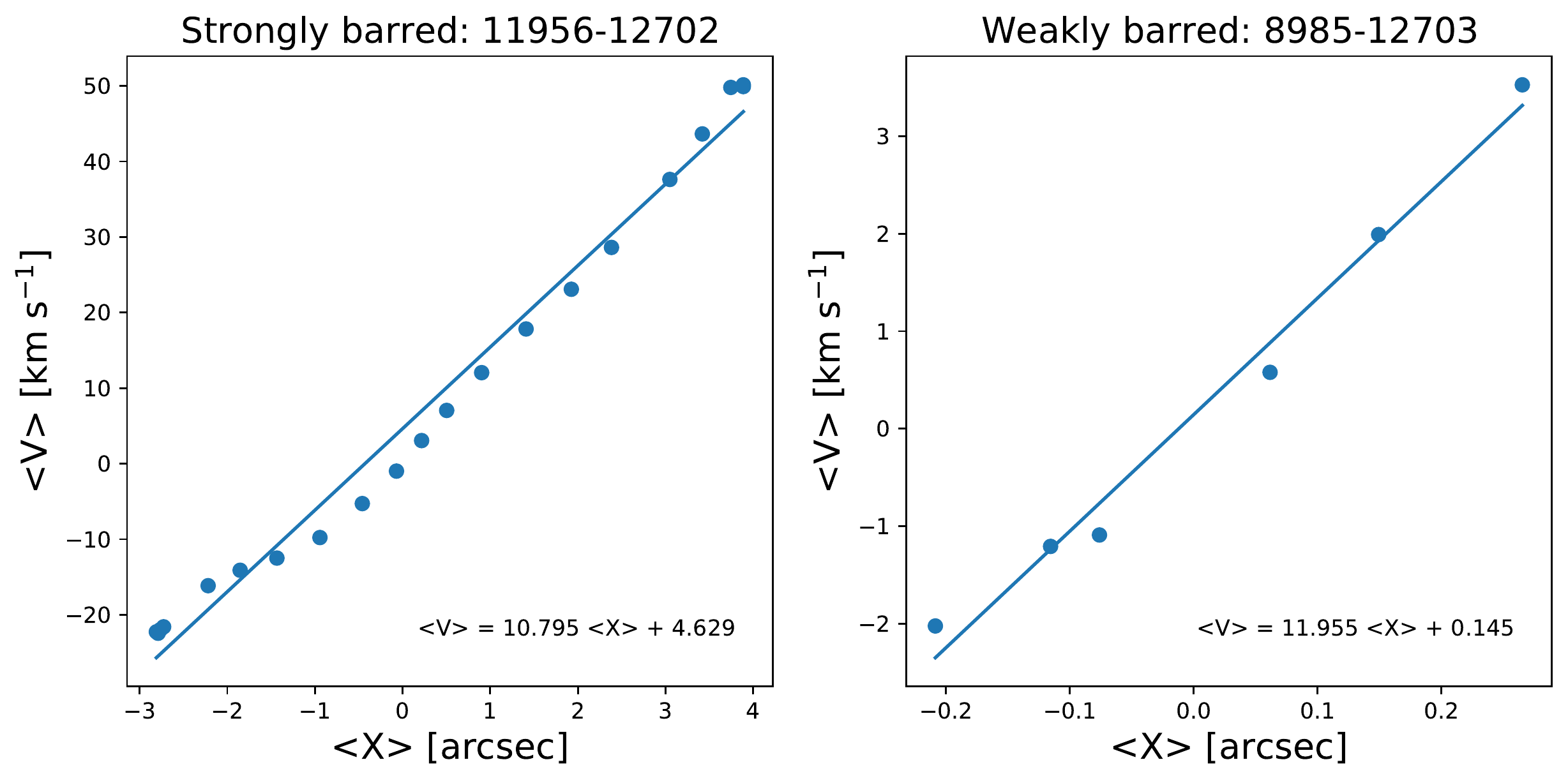}
    \caption{The kinematic integral, $\left<V\right>$, is plotted against the photometric integral, $\left<X\right>$, for all the pseudo-slits for a strongly barred galaxy (left) and a weakly barred galaxy (right). The equation of the best-fit line going through these points is shown in each plot. The slope of these lines is equal to $\Omega_{\rm b} \sin{\left( i \right)}$.}
    \label{fig:example_X_V}
\end{figure}

We do not calculate the ratio of $\left<V\right>$ and $\left<X\right>$ directly. Instead, we plot $\left<V\right>$ against $\left<X\right>$ for the different pseudo-slits and the slope of the best-fit line going through these points will then be equal to $\Omega_{\rm b} \sin{\left( i \right)}$. This is done to help avoid centering errors and account for incorrect estimates of the systemic velocity \citep{guo_2019}. An example of such a plot can be found in Figure \ref{fig:example_X_V}.

It is important to note that for an axisymmetric disc, the weighted mean position and velocity integrals will equal zero. This means that any non-zero values will be due to additional structures such as the bar (if it is not aligned or perpendicular to the LON).

The code used in this work to calculate the bar pattern speed is publicly available \href{https://doi.org/10.5281/zenodo.7567945}{here}\footnote{\url{https://doi.org/10.5281/zenodo.7567945}}.

\subsection{Concerns and limitations}
\label{sec:tw_limits}

As alluded to in Section \ref{sec:tw_theory}, the TW method requires that the chosen tracer satisfies the continuity equation. Multiple studies have used gas as the tracer and have been successful in determining the pattern speed using the TW method \citep{zimmer_2004,hernandez_2005,emsellem_2006,fathi_2009,gabbasov_2009}.

Many studies have also successfully determined pattern speeds by using stars as the tracer \citep{merrifield_1995,debattista_2002,aguerri_2003,corsini_2007}, although they usually limited their sample to early-type barred galaxies. This is because there were concerns that dust obscuration and star formation in late-type galaxies could cause the surface brightness to not trace the mass distribution properly. However, other papers show that it is possible to use the TW method on late-type galaxies, despite initial concerns \citep{gerssen_2003,gerssen_2007,treuthardt_2007,aguerri_2015,cuomo_2019,guo_2019,garma_oehmichen_2020}. More recently, \citet{williams_2021} applied the TW method to stellar and gaseous tracers (using both CO and H$\alpha$) and found significantly different results. They attributed this inconsistency to the clumpy nature of the gaseous tracers they used, which resulted in incorrect pattern speed measurements. Thus, in this work we decided to use stars as our tracer. 

As \citet{garma_oehmichen_2020} show, centering issues are not negligible and it is crucial that the LON goes through the centre of the galaxy. In this work, we find the centre by smoothing the stellar flux data with a Gaussian filter and finding the brightest pixel in the smoothened data.

The TW method is also only applicable to galaxies with regular kinematics and on galaxies with intermediate inclinations (\todo{20}$^{\circ}$ $< i < $ \todo{70}$^{\circ}$) \citep{tremaine_1984,aguerri_2015,cuomo_2019,garma_oehmichen_2020}. Edge-on galaxies do not have enough spatial data, while the stellar velocity is not well constrained in face-on galaxies. Additionally, detecting bars in edge-on galaxies is very difficult and unreliable. It is known that the TW method is very sensitive to incorrect estimates of the PA of the galaxy \citep{debattista_2003,zou_2019,garma_oehmichen_2020}. Thus, a correct and reliable estimate of the position angle is crucial. We try to account for this sensitivity by performing a Monte Carlo (MC) simulation over the uncertainty of the PA (see Section \ref{sec:inputs} for more details). For the TW method to work, it is also important that the bar is not aligned with the major or minor axis of the galaxy, as otherwise the integrals will cancel out. We also need to be able to place a sufficient amount of slits, otherwise the straight line in the $\left<V\right>$ over $\left<X\right>$ plot that is used to determine $\Omega_{\rm b}$ is not well constrained. As the slits have to be placed on top of the bar, the TW method is not ideal for the shortest of bars, where only very few slits can be placed. This will disproportionally affect weak bars, which should be kept in mind. The specific thresholds we impose are detailed in Section \ref{sec:sample_selection}.

\subsection{Calculation of corotation radius and $\mathcal{R}$}
\label{sec:calc_rcr_r}

After the bar pattern speed is obtained using the TW method, one can calculate the corotation radius ($R_{\rm CR}$). This is where the centrifugal and gravitational forces balance each other in the rest frame of the bar, which means that the stars in the disc will have the same angular velocity as the bar pattern speed at the corotation radius \citep{cuomo_2019,guo_2019}. Various papers calculate this by doing $R_{\rm CR} = V_{\rm c} / \Omega_{\rm b}$, where $V_{\rm c}$ is the circular velocity in the flat part of the rotation curve \citep{aguerri_2015, cuomo_2019, guo_2019}. However, this assumes that the corotation radius lies in the region where the rotation curve has flattened. This is not necessarily the case and can lead to incorrect estimates of $R_{\rm CR}$ and $\mathcal{R}$. 

\begin{figure}
	\includegraphics[width=\columnwidth]{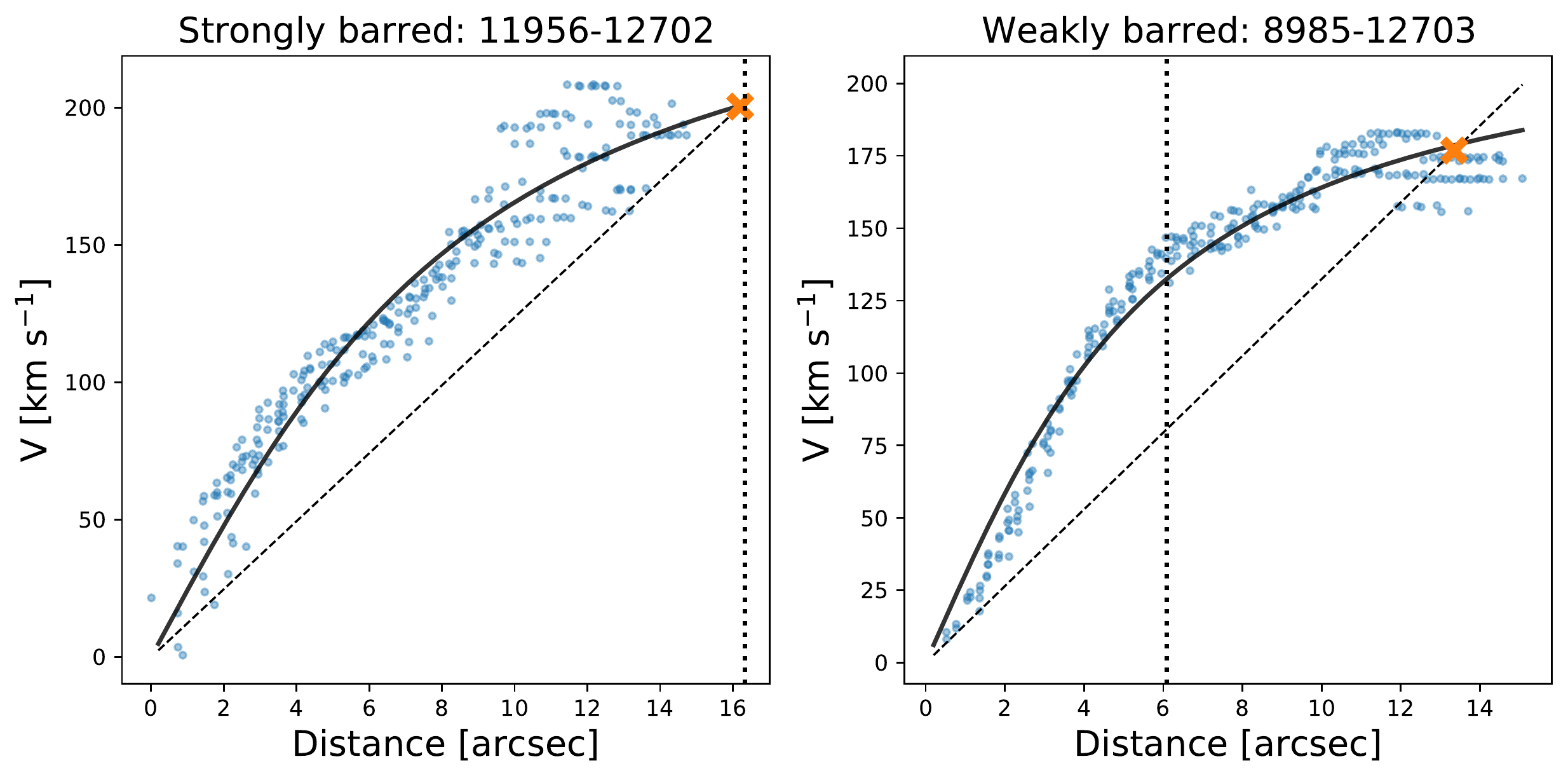}
    \caption{Visualisation of how the corotation radius is obtained, for a strongly barred galaxy (left) and a weakly barred galaxy (right). The blue dots are stellar velocity measurement from MaNGA in a \todo{5} arcsec aperture along the major axis of the galaxy. The black line is the best-fit rotation curve, please refer to Section \ref{sec:rotation_curve} for more details on how the rotation curve is calculated. The black dashed line is obtained by multiplying $\Omega_{\rm b}$ with a radius range. The distance where this line and the rotation curve intersect (indicated by the orange cross), defines the corotation radius. The dotted vertical line is the deprojected bar radius.}
    \label{fig:example_velcurve}
\end{figure}

Instead, we will use the rotation curve of the galaxy to calculate the corotation radius. The rotation curve can be obtained by using the MaNGA stellar velocity data (see Section \ref{sec:rotation_curve}). The bar pattern speed is multiplied by a radius range, which effectively indicates how fast the tracer moves at any radius for that particular pattern speed. The radius at which this curve intersects with the galaxy rotation curve, is the corotation radius. An example of this can be found in Figure \ref{fig:example_velcurve}. 

\citet{garma_oehmichen_2020} used a similar approach to ours and compared their results to results obtained by using the $R_{\rm CR} = V_{\rm c} / \Omega_{\rm b}$ method. On average, they found a relative difference of $\sim$15\%, indicating that the simplified approach introduces a significant bias. 

\begin{figure*}
	\includegraphics[width=\textwidth]{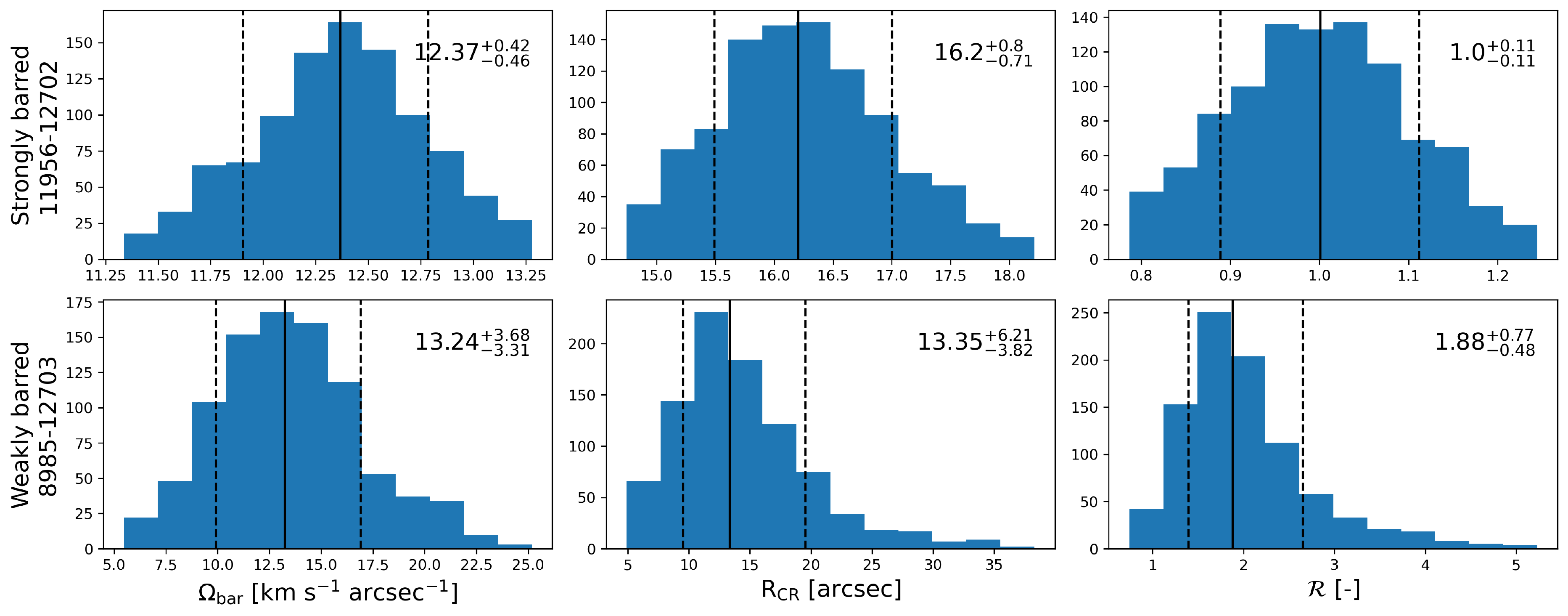}
    \caption{The posterior distributions of the pattern speed (left column), corotation radius (middle column) and $\mathcal{R}$ (right column) for a strongly barred galaxy (top row) and weakly barred galaxy (bottom row), obtained from performing Monte Carlo (MC) simulations of a 1,000 iterations in order to characterise the uncertainty on each measurement. The median value is indicated in every plot by a black vertical line, while the 16$^{\rm th}$ and 84$^{\rm th}$ percentile are shown by the dashed vertical lines. These values are also printed in each subplot.}
    \label{fig:example_MCs}
\end{figure*}

The corotation radius can be used to calculate the dimensionless parameter $\mathcal{R}$, defined as $\mathcal{R} = R_{\rm CR} / R_{\rm bar}$, where $R_{\rm bar}$ is the deprojected bar radius. We obtain estimates for the uncertainty on $\Omega_{\rm b}$, $R_{\rm CR}$ and $\mathcal{R}$ by performing a Monte Carlo simulation using the errors on the input variables and assuming Gaussianity (see Section \ref{sec:inputs} for more details). The posterior distributions of the final pattern speed, corotation radius and $\mathcal{R}$ for our example galaxies are shown in Figure \ref{fig:example_MCs}.

$\mathcal{R}$ can be used to classify bars into fast ($1.0 < \mathcal{R} < 1.4$) and slow ($\mathcal{R} > 1.4$) bars \citep{debattista_2000,rautiainen_2008,aguerri_2015}. Thus, slow bars have bar lengths that are shorter that the corotation radius, whereas fast bars end near the corotation radius. It is suggested that bars cannot extend beyond corotation \citep{contopoulos_1980,contopoulos_1981, athanassoula_1992}, which means that bars with $\mathcal{R} < 1.0$ are not expected. However, multiple studies have observed these so-called ultrafast bars \citep{buta_2009, aguerri_2015, cuomo_2019,guo_2019,garma_oehmichen_2020}.


\section{Data}
\label{sec:data}

\subsection{MaNGA survey}

We need resolved stellar velocity and stellar flux data in order to implement the TW method, which we obtain from the Mapping Nearby Galaxies at Apache Point Observatory (MaNGA) survey \citep{bundy_2015}. MaNGA is part of the Sloan Digital Sky Survey IV (SDSS-IV) collaboration \citep{blanton_2017}. More specifically, we used data from the seventeenth data release of SDSS \citep{abdurrouf_2022}. MaNGA used the Baryon Oscillation Spectroscopic Survey (BOSS) Spectrograph, which has a resolution of R $\sim 2000$ and a wavelength coverage of 3600 - 10,000 \AA\ \citep{smee_2013}, on the 2.5m Sloan Telescope at Apache Point Observatory \citep{gunn_2006}. Every integral field unit (IFU) consists of 19-127 optical fibers, stacked hexagonally \citep{drory_2015}. Most galaxies are covered out to 1.5 effective radii (R$_{e}$), while a third are covered out to 2.5 R$_{e}$. We make use of the maps that are binned to S/N $\sim$ 10 using the Voronoi binning algorithm \citep{westfall_2019}. For more information on the observing strategy, survey design, data reduction process, sample selection and the data analysis pipeline, please refer to \citet{law_2015,yan_2016,law_2016,wake_2017, belfiore_2019, westfall_2019}. All the stellar masses and SFRs in this paper come from the Pipe3D value added catalog \citep{sanchez_2016,sanchez_2016b}. The SFRs in Pipe3D are estimated from the H$\alpha$ flux and is dust and aperture corrected. For more details, please refer to \citep{sanchez_2016b}. Finally, this paper made extensive use of the \textit{Marvin} software in order to access MaNGA data \citep{cherinka_2019}.

\subsection{Galaxy Zoo and the Legacy Survey}
\begin{figure}
	\includegraphics[width=\columnwidth]{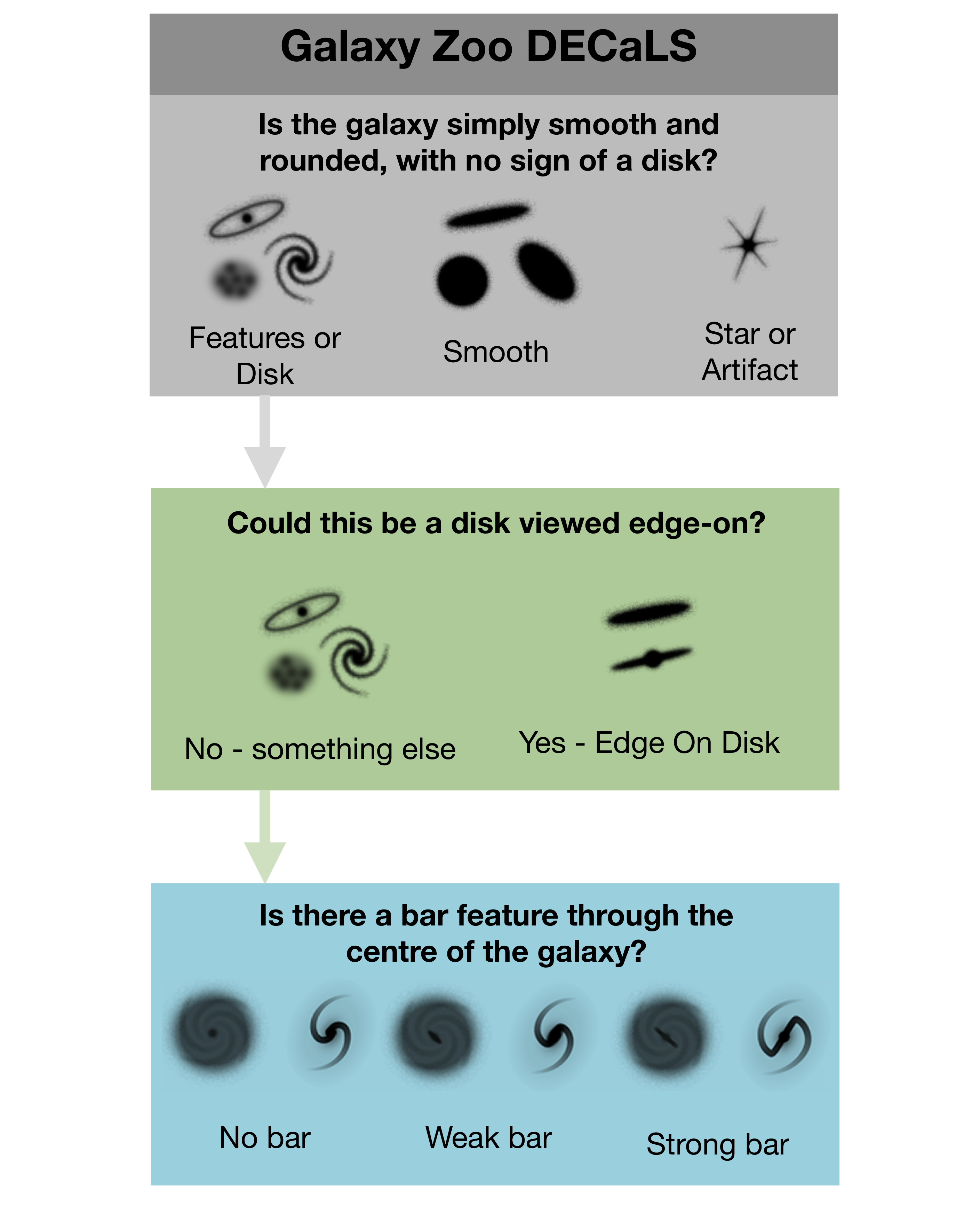}
    \caption{The decision tree of GZ DESI up to the bar question. It is worth noting that volunteers will only reach the bar question after they said the target is a disk galaxy that is not viewed edge-on. The full decision tree is shown in \citet{walmsley_2022}.}
    \label{fig:GZ_tree}
\end{figure}

We have used the Galaxy Zoo (GZ) project to obtain morphological classifications and find weak and strong bars. Here, citizen scientists classify galaxies according to a decision tree \citep{lintott_2008, lintott_2011}. We made use of the latest iteration of Galaxy Zoo, namely Galaxy Zoo DESI (GZ DESI, \inprep{Walmsley et al., in prep.}). GZ DESI sources images from the DESI Legacy Imaging Surveys\footnote{\url{www.legacysurvey.org/}} \citep{dey_2019}, which consists of three individual projects: the Dark Energy Camera Legacy Survey (DECaLS), the Beijing-Arizona Sky Survey (BASS) and the Mayall z-band Legacy Survey (MzLS), which covers $\sim14,000$ deg$^{2}$ of sky. As shown by \citet{geron_2021}, the DESI Legacy Imaging Surveys are sufficiently deep so that weak bars are visible and can be identified by the volunteers (the median $5\sigma$ point source depth of DECaLS is r = 23.6, \citealp{dey_2019}). GZ DESI uses classifications from citizen scientists to train machine classifications based on the Bayesian convolutional neural networks described in \citet{walmsley_2022}, which we rely on for our morphology measurements. The decision tree of GZ DESI, up to the bar question, is shown in Figure \ref{fig:GZ_tree}. Note that volunteers will only reach the bar question after they identified the target as being a disk that is not edge-on.

\subsection{Inclination, bar length and position angles}
\label{sec:inputs}

In order to perform the TW method, we need multiple additional parameters. We need the inclination of the galaxy, the kinematic position angle of the galaxy, the length of the bar and the position angle of the bar.

\subsubsection{Position angle}

Various papers using the TW method often use photometric approaches to obtain the position angle of the galaxy, such as fitting ellipses in the sky plane \citep{aguerri_2015, cuomo_2019,guo_2019}. The outermost isophotes are then used to estimate the position angle. However, multiple issues are associated with this method. First of all, there is no systematic way to determine which and how many outer isophotes to use to determine the position angle. Secondly, the presence of a bar, especially a strong bar, will influence these estimates. Additionally, spiral arms, rings, companion galaxies and foreground stars will all also affect these measurements significantly. 

However, MaNGA allows us to use kinematic position angles rather than photometric ones. We obtained the global kinematic position angle and its uncertainty by using the Python package \texttt{PaFit} on the stellar velocity maps from MaNGA. Small-scale disturbances in the velocity field are removed by smoothing the stellar velocity maps using a 10x10 pixel sliding window that calculates the median at every position, before the kinematic position angle is calculated. This package is based on the method detailed in Appendix C of \citet{krajnovic_2006}\footnote{\url{www-astro.physics.ox.ac.uk/\~cappellari/software\#pafit}}. It constructs a bi-anti-symmetric map based on the original input. The position angle that minimises the difference between the original velocity map and the symmetrised map is considered to be the best-fit global kinematic position angle. The error on the best-fit position angle is defined as the range of angles for which the difference in $\chi^{2}$ with the best-fit angle is less than 9 \citep{krajnovic_2006}, which corresponds to a 3$\sigma$ confidence limit. Multiple other papers have successfully used this code to study galaxy kinematics before \citep{cappellari_2007,krajnovic_2011}. The kinematic position angle does not suffer from the issues that plague the photometric position angle, which is why we used the kinematic one in this work. However, the kinematic position angle is not infallible. The bar can twist the inner parts of the disc velocity field, which can affect the measurement of the kinematic position angle. Additionally, the coverage of the IFU, inclination of the galaxy and the difference in position angle between the bar and disc will have some influence as well. These effects are described in more detail in Appendix A3 of \citet{guo_2019}.

\subsubsection{Inclination}
\label{sec:inclination}

Estimating the inclination of a barred galaxy, especially a strongly barred one, is not straight forward. As a strong bar is a very obvious component, it will make the galaxy appear more inclined. Therefore, we carefully measured the inclinations of our barred galaxies ourselves, using the elliptical isophote analysis technique described by \citet{jedrzejewski_1987} using the Python package \texttt{photutils}\footnote{\url{https://photutils.readthedocs.io/en/stable/index.html}} on the r-band images from the Legacy Survey. We typically averaged the ellipticity of the outermost 5\% of fitted isophotes, which usually corresponded to 5 isophotes. However, to guarantee the bar does not affect our measurement, we excluded any isophotes that are within the bar region. This meant that we used less than 5 isophotes for some targets that had long bars. The ellipticity profiles and r-band images of all our targets were inspected individually to make sure that the final value was correct. However, spiral arms, foreground stars and rings will all bias this measurement to some degree. To estimate the error on this ellipticity measurement, we correctly combine the errors associated with the isophotes used to calculate the ellipticity.

\subsubsection{Bar length and bar position angle}

There are multiple ways to determine the length of the bar. One possibility involves ellipse fitting again \citep{laine_2002, erwin_2005, marinova_2007, aguerri_2009}, but as bars are associated with spiral arms, rings and ansae, this method is prone to inconsistencies. Other methods include Fourier decomposition \citep{aguerri_2000} and using explainable artificial intelligence and saliency mapping techniques \citep{bhambra_2022}. Additionally, the Galaxy Zoo:3D project provides bar masks for galaxies in MaNGA, based on SDSS images, which can be used to estimate bar length \citep{masters_2021}. 

A more straight-forward approach is to manually measure bar lengths. Manual bar length measurements have been successfully used in various studies \citep{erwin_2019, geron_2021}. Additionally, \citet{hoyle_2011} have found that manual bar length measurements between different volunteers agree within 10\% of each other and they show that manual measurements can be unbiased and robust against systematic effects. Additionally, \citet{diaz_garcia_2016} have found that their manual bar lengths agree with bar lengths determined by various other automated techniques.

Thus, manual bar length measurements are used in this work. The bar lengths were measured manually by one of the authors (TG) on \textit{grz}-images obtained from the Legacy Survey. A number of measures were put in place to make sure these measurements were done as consistently and correctly as possible. For example, the order of the measurements was completely randomised, so that it was was not known whether the bar that was being measured was classified as a strong or weak bar by GZ. The measurements themselves were done in DS9 \citep{joye_2003} with a measurement tool that automatically records the distance measured. Additionally, every bar was measured twice and the final bar length distribution is modelled by a Gaussian centered around the average of the two measurements and with an uncertainty equal to half the difference between the two measurements. Finally, all measurements were inspected again afterwards to make sure no mistakes were made. These bar lengths were successfully used before in \citet{geron_2021}, where they were compared to another bar length catalog \citep{hoyle_2011}. The bar lengths are deprojected using the method described by \citet{gadotti_2007}:

\begin{equation}
    R_{\rm b, deproj} = R_{\rm b, obs} \sqrt{ \cos^{2}{\phi} + \sin^{2}{\phi} / \cos^{2}{i}}\;,
    \label{eq:deprojection}
\end{equation}

where $i$ is the inclination of the galaxy, $\phi$ is the difference between the position angle of the bar and of the galaxy and R$_{\rm b, obs}$ and R$_{\rm b, deproj}$ are the observed and deprojected bar lengths, respectively. The position angles of the bar were also obtained from these manual measurements and are similarly modelled by a Gaussian centered around the average of the measurements and with an uncertainty equal to half the difference between the two measurements.

The uncertainties on the inclination, disk PA, bar length and bar PA are used to estimate the uncertainty on the bar pattern speed, corotation radius and $\mathcal{R}$. This is achieved by assuming Gaussianity over these input parameters and performing a Monte Carlo simulation with 1,000 iterations.

An overview of all the input parameters for 50 randomly selected targets is given in Table \ref{tab:input_params}. The full table can be found online \href{https://doi.org/10.5281/zenodo.7567945}{here}\footnote{\url{https://doi.org/10.5281/zenodo.7567945}}.

\begin{table*}
    \centering
    \caption{The plate-ifu number, right ascension, declination, inclination, position angle of the disk, position angle of the bar, the (projected) bar radius, redshift and the bar type for 50 randomly selected galaxies. The full table can be found online \href{https://doi.org/10.5281/zenodo.7567945}{here}.}

    \begin{tabular}{llllllllll}
        \hline
        \hline
        Plate-ifu & RA [$^{\circ}$] & DEC [$^{\circ}$] & Inclination [$^{\circ}$] & PA$_{\rm disk}$ [$^{\circ}$] & PA$_{\rm bar}$ [$^{\circ}$] & R$_{\rm bar}$ [arcsec] & R$_{\rm bar, deproj}$ [kpc] & Redshift & Bar type \\
        \hline
        11014-12705 & 194.4765 & 27.4906 & 42.07$\pm$0.47 & 85.10$\pm$1.07 & 59.22$\pm$40.72 & 8.16$\pm$0.26 & 3.06$\pm$0.10 & 0.0166 & Strong bar \\[0.09cm]
        9504-3704 & 123.3717 & 29.0372 & 37.22$\pm$6.03 & 168.50$\pm$1.40 & 104.64$\pm$68.15 & 3.74$\pm$0.32 & 4.40$\pm$0.38 & 0.0479 & Strong bar \\[0.09cm]
        8245-12702 & 136.1968 & 22.0285 & 60.15$\pm$0.26 & 14.80$\pm$0.42 & 164.79$\pm$89.37 & 10.56$\pm$0.25 & 8.49$\pm$0.20 & 0.0343 & Weak bar \\[0.09cm]
        11979-12703 & 252.9135 & 23.9723 & 36.58$\pm$1.70 & 116.40$\pm$0.83 & 149.08$\pm$177.18 & 5.71$\pm$0.01 & 4.22$\pm$0.00 & 0.0356 & Weak bar \\[0.09cm]
        9027-12704 & 245.3466 & 32.3490 & 54.90$\pm$1.87 & 134.30$\pm$1.55 & 87.73$\pm$31.14 & 10.57$\pm$0.28 & 9.73$\pm$0.26 & 0.0347 & Strong bar \\[0.09cm]
        8079-9101 & 42.8963 & -0.7338 & 47.22$\pm$0.21 & 13.90$\pm$0.82 & 44.72$\pm$15.30 & 8.08$\pm$0.03 & 4.34$\pm$0.02 & 0.0232 & Strong bar \\[0.09cm]
        8624-9102 & 263.8926 & 59.8899 & 44.27$\pm$6.58 & 139.30$\pm$0.87 & 104.74$\pm$100.54 & 4.81$\pm$0.20 & 3.23$\pm$0.14 & 0.0284 & Strong bar \\[0.09cm]
        10220-9101 & 120.8259 & 31.7764 & 39.91$\pm$1.08 & 62.80$\pm$0.75 & 82.59$\pm$82.09 & 5.95$\pm$0.06 & 4.50$\pm$0.04 & 0.0364 & Strong bar \\[0.09cm]
        11979-9101 & 252.3505 & 22.9414 & 32.97$\pm$1.12 & 2.10$\pm$0.90 & 39.90$\pm$219.31 & 6.55$\pm$0.44 & 6.27$\pm$0.42 & 0.0443 & Weak bar \\[0.09cm]
        8723-12701 & 126.9739 & 55.1586 & 62.95$\pm$1.73 & 1.50$\pm$0.45 & 37.38$\pm$10.45 & 3.93$\pm$0.18 & 4.06$\pm$0.19 & 0.0388 & Weak bar \\[0.09cm]
        9881-12704 & 205.3262 & 24.4962 & 46.05$\pm$0.89 & 68.10$\pm$1.37 & 94.43$\pm$40.61 & 11.45$\pm$0.38 & 6.69$\pm$0.22 & 0.0269 & Strong bar \\[0.09cm]
        11868-12703 & 248.7371 & 25.6926 & 60.00$\pm$2.06 & 116.00$\pm$0.92 & 47.18$\pm$87.40 & 9.11$\pm$0.17 & 9.91$\pm$0.19 & 0.0436 & Strong bar \\[0.09cm]
        11965-9102 & 231.5925 & 9.3964 & 50.16$\pm$0.48 & 104.20$\pm$0.97 & 126.51$\pm$124.65 & 5.03$\pm$0.18 & 3.78$\pm$0.14 & 0.0323 & Strong bar \\[0.09cm]
        10492-12702 & 124.0641 & 57.5305 & 40.46$\pm$1.89 & 178.70$\pm$0.55 & 119.04$\pm$11.38 & 3.87$\pm$0.21 & 2.67$\pm$0.15 & 0.0272 & Strong bar \\[0.09cm]
        8257-3703 & 166.6557 & 46.0388 & 61.50$\pm$1.07 & 155.30$\pm$1.07 & 122.14$\pm$2.98 & 4.61$\pm$0.06 & 2.74$\pm$0.03 & 0.0250 & Strong bar \\[0.09cm]
        8602-12705 & 247.4627 & 39.7665 & 41.28$\pm$1.10 & 145.30$\pm$0.90 & 7.93$\pm$50.26 & 10.74$\pm$0.45 & 7.83$\pm$0.33 & 0.0318 & Strong bar \\[0.09cm]
        8323-12705 & 196.7939 & 34.2980 & 64.63$\pm$0.28 & 130.30$\pm$0.48 & 118.90$\pm$39.45 & 11.06$\pm$0.27 & 8.01$\pm$0.20 & 0.0338 & Weak bar \\[0.09cm]
        8442-9102 & 200.2228 & 32.1908 & 34.07$\pm$7.31 & 15.70$\pm$1.57 & 90.08$\pm$16.89 & 6.20$\pm$0.29 & 4.23$\pm$0.20 & 0.0230 & Strong bar \\[0.09cm]
        10492-6102 & 121.9606 & 56.6935 & 54.47$\pm$1.07 & 6.80$\pm$0.58 & 69.65$\pm$149.39 & 3.72$\pm$0.33 & 3.14$\pm$0.28 & 0.0297 & Weak bar \\[0.09cm]
        11956-12702 & 187.7783 & 52.4143 & 60.81$\pm$1.05 & 53.00$\pm$0.55 & 75.56$\pm$5.16 & 13.46$\pm$0.49 & 12.74$\pm$0.46 & 0.0400 & Strong bar \\[0.09cm]
        10226-3704 & 37.7793 & -1.1052 & 26.27$\pm$0.58 & 77.30$\pm$1.48 & 130.76$\pm$143.22 & 5.07$\pm$0.13 & 4.68$\pm$0.12 & 0.0406 & Strong bar \\[0.09cm]
        8145-3704 & 117.5703 & 27.8570 & 51.56$\pm$0.45 & 90.00$\pm$0.87 & 48.80$\pm$9.61 & 4.90$\pm$0.19 & 3.25$\pm$0.12 & 0.0275 & Strong bar \\[0.09cm]
        8456-6101 & 151.2209 & 44.6361 & 32.71$\pm$0.66 & 129.70$\pm$1.48 & 73.54$\pm$105.52 & 9.07$\pm$0.27 & 5.20$\pm$0.16 & 0.0232 & Strong bar \\[0.09cm]
        12651-9101 & 250.6869 & 26.5976 & 58.77$\pm$0.35 & 175.00$\pm$0.60 & 13.44$\pm$13.03 & 11.81$\pm$0.24 & 11.41$\pm$0.23 & 0.0451 & Strong bar \\[0.09cm]
        8324-9101 & 197.4380 & 45.9127 & 50.91$\pm$1.95 & 101.70$\pm$0.72 & 146.15$\pm$405.97 & 2.68$\pm$0.25 & 1.88$\pm$0.18 & 0.0288 & Weak bar \\[0.09cm]
        8084-6101 & 51.6942 & -0.6482 & 51.44$\pm$0.83 & 64.50$\pm$1.07 & 8.92$\pm$26.48 & 4.84$\pm$0.65 & 2.77$\pm$0.37 & 0.0205 & Strong bar \\[0.09cm]
        11004-12704 & 197.0588 & 27.5159 & 29.24$\pm$0.83 & 75.70$\pm$1.42 & 143.22$\pm$45.98 & 7.18$\pm$0.15 & 3.93$\pm$0.08 & 0.0244 & Weak bar \\[0.09cm]
        9886-12701 & 236.3470 & 24.5068 & 53.22$\pm$0.91 & 91.60$\pm$0.88 & 152.87$\pm$51.34 & 3.78$\pm$0.11 & 2.21$\pm$0.06 & 0.0230 & Weak bar \\[0.09cm]
        8602-12701 & 247.0482 & 39.8219 & 39.32$\pm$5.72 & 156.40$\pm$0.60 & 16.65$\pm$91.89 & 9.97$\pm$0.34 & 6.12$\pm$0.21 & 0.0268 & Strong bar \\[0.09cm]
        9190-12703 & 54.4953 & -6.2706 & 54.41$\pm$0.45 & 49.60$\pm$0.57 & 116.52$\pm$29.00 & 5.75$\pm$0.44 & 3.46$\pm$0.26 & 0.0221 & Weak bar \\[0.09cm]
        8978-9101 & 247.9080 & 41.4936 & 32.99$\pm$0.42 & 95.50$\pm$1.07 & 78.40$\pm$59.71 & 5.22$\pm$0.38 & 3.23$\pm$0.23 & 0.0303 & Weak bar \\[0.09cm]
        8137-9102 & 117.0386 & 43.5907 & 46.10$\pm$0.31 & 132.80$\pm$0.82 & 116.44$\pm$111.81 & 7.45$\pm$1.41 & 4.98$\pm$0.94 & 0.0311 & Weak bar \\[0.09cm]
        8596-12702 & 230.1723 & 49.1065 & 47.89$\pm$0.36 & 33.20$\pm$0.50 & 94.12$\pm$154.39 & 3.47$\pm$0.15 & 3.53$\pm$0.15 & 0.0383 & Strong bar \\[0.09cm]
        12495-6102 & 160.4608 & 4.3308 & 64.23$\pm$0.21 & 48.20$\pm$0.73 & 85.44$\pm$120.89 & 4.05$\pm$0.13 & 3.05$\pm$0.10 & 0.0268 & Weak bar \\[0.09cm]
        9028-12702 & 242.9751 & 30.3328 & 49.76$\pm$0.51 & 137.60$\pm$0.57 & 174.52$\pm$9.26 & 6.18$\pm$0.06 & 4.23$\pm$0.04 & 0.0301 & Weak bar \\[0.09cm]
        8619-12701 & 322.2428 & 11.3665 & 23.35$\pm$1.32 & 83.10$\pm$0.97 & 99.10$\pm$45.60 & 10.78$\pm$0.42 & 6.57$\pm$0.26 & 0.0292 & Strong bar \\[0.09cm]
        11834-12705 & 223.3961 & 0.0104 & 56.12$\pm$0.43 & 86.80$\pm$0.52 & 68.02$\pm$10.16 & 6.48$\pm$0.61 & 5.88$\pm$0.55 & 0.0424 & Strong bar \\[0.09cm]
        8982-6104 & 203.0571 & 26.9500 & 57.59$\pm$4.18 & 155.70$\pm$0.68 & 1.95$\pm$38.77 & 3.14$\pm$0.08 & 2.66$\pm$0.07 & 0.0353 & Strong bar \\[0.09cm]
        7962-12704 & 260.8831 & 27.4587 & 53.11$\pm$5.55 & 84.90$\pm$1.02 & 116.02$\pm$360.05 & 3.12$\pm$0.36 & 1.52$\pm$0.17 & 0.0223 & Weak bar \\[0.09cm]
        9095-9102 & 243.0849 & 23.0020 & 51.93$\pm$1.77 & 80.30$\pm$0.75 & 27.62$\pm$92.82 & 2.86$\pm$0.07 & 2.29$\pm$0.06 & 0.0323 & Weak bar \\[0.09cm]
        8947-9101 & 168.7345 & 50.3349 & 42.67$\pm$0.77 & 24.20$\pm$1.60 & 159.58$\pm$432.76 & 3.02$\pm$0.38 & 3.12$\pm$0.39 & 0.0471 & Strong bar \\[0.09cm]
        9089-12704 & 241.1484 & 25.1899 & 62.55$\pm$0.38 & 22.10$\pm$0.43 & 33.23$\pm$363.90 & 10.77$\pm$1.84 & 7.37$\pm$1.26 & 0.0318 & Strong bar \\[0.09cm]
        11976-12705 & 243.5467 & 19.3149 & 37.91$\pm$2.27 & 57.00$\pm$1.22 & 0.23$\pm$250.30 & 2.83$\pm$0.18 & 1.95$\pm$0.12 & 0.0308 & Weak bar \\[0.09cm]
        11743-6104 & 118.8186 & 14.4344 & 54.33$\pm$1.64 & 98.50$\pm$0.97 & 83.24$\pm$69.38 & 5.51$\pm$0.20 & 3.44$\pm$0.12 & 0.0291 & Weak bar \\[0.09cm]
        10492-9101 & 121.8889 & 56.4257 & 48.75$\pm$1.52 & 45.00$\pm$0.17 & 13.29$\pm$358.61 & 4.71$\pm$0.10 & 2.99$\pm$0.06 & 0.0268 & Strong bar \\[0.09cm]
        8244-3702 & 131.8150 & 51.2458 & 37.94$\pm$1.07 & 68.10$\pm$1.97 & 172.74$\pm$168.22 & 2.99$\pm$0.28 & 2.05$\pm$0.19 & 0.0275 & Strong bar \\[0.09cm]
        11834-6103 & 223.7898 & 0.7816 & 40.00$\pm$5.57 & 2.60$\pm$1.55 & 25.79$\pm$89.39 & 2.45$\pm$0.06 & 2.27$\pm$0.06 & 0.0430 & Strong bar \\[0.09cm]
        8621-12704 & 351.9572 & 15.1192 & 37.83$\pm$1.45 & 145.80$\pm$0.53 & 175.14$\pm$34.39 & 8.57$\pm$0.35 & 7.98$\pm$0.33 & 0.0419 & Strong bar \\[0.09cm]
        8615-6104 & 319.9019 & 0.7042 & 52.71$\pm$0.22 & 170.00$\pm$0.57 & 121.55$\pm$170.89 & 6.62$\pm$0.23 & 5.62$\pm$0.20 & 0.0347 & Weak bar \\[0.09cm]
        12622-9102 & 200.6963 & 32.6233 & 69.45$\pm$1.04 & 97.20$\pm$0.48 & 28.84$\pm$59.63 & 2.43$\pm$0.55 & 2.86$\pm$0.65 & 0.0426 & Weak bar \\[0.09cm]
        and \todo{175} more rows...\\
        \hline
    \end{tabular}

    \label{tab:input_params}
\end{table*}

\subsection{Rotation curve}
\label{sec:rotation_curve}

As mentioned in Section \ref{sec:calc_rcr_r}, we need the rotation curve of the galaxy in order to obtain the corotation radius. The rotation curve can be determined from the stellar velocity IFU data from MaNGA. We look at the spaxels in a \todo{5} arcsec aperture along the position angle of the galaxy. The true stellar velocity in every spaxel is calculated from the observed stellar velocity by doing:

\begin{equation}
    V_{\rm rot} = V_{\rm obs} / \left( \sin{i} \times \cos{\phi} \right)\;,
\end{equation}

where V$_{\rm obs}$ and V$_{\rm rot}$ are the observed and true velocity in that spaxel, $i$ is the inclination of the galaxy and $\phi$ is the azimuthal angle measured relative to the position angle of the galaxy. The distance to the centre of the galaxy is deprojected using equation \ref{eq:deprojection}. The corrected velocities and deprojected distances are used to fit a two parameter arctan function, described in \citet{courteau_1997}:

\begin{equation}
    V_{\rm rot} = V_{\rm sys} + \frac{2}{\pi} V_{\rm c} \arctan{\left(\frac{r - r_{0}}{r_{t}}\right)}\;,
\end{equation}

where V$_{\rm sys}$ is the systemic velocity, V$_{\rm c}$ is the asymptotic velocity, r$_{0}$ is the spatial centre of the galaxy and r$_{t}$ is the transition radius. The rotation curve flattens at r$_{t}$ and goes towards V$_{\rm c}$ in this model. For our purposes, V$_{\rm sys}$ and r$_{0}$ are assumed to equal zero.

\subsection{Sample selection}
\label{sec:sample_selection}

We use the machine classifications from GZ DESI (\inprep{Walmsley et al. in prep.}). GZ works based on a decision tree structure. As you can see in Figure \ref{fig:GZ_tree}, this means that the question ``\textit{Is there a bar feature through the centre of the galaxy?}'' is only answered when the galaxy is a not edge-on disk galaxy. To guarantee reliable bar classifications, we must apply additional thresholds on the fraction of people that would have been asked the bar question (N$_{\rm bar}$\footnote{Please note that, as we are using machine classifications, in this context, $N_{\textrm{bar}}$ is not the amount of people that have been asked the bar question. Rather, it is the estimated fraction of people that would have been asked the bar question.}) and the fraction of people that would have voted for a certain answer (e.g., $p_{\rm strong\;bar}$), as predicted by the automated classifications. We choose to apply $p_{\textrm{features/disk}} \geq 0.27$, $p_{\textrm{not edge-on}} \geq 0.68$ and $N_{\textrm{bar}}\geq 0.5$. For more information on these thresholds, please refer to \citet{geron_2021} and \citet{walmsley_2022}. These thresholds resulted in a sample of \todo{3,106} galaxies that consists of relatively face-on disk galaxies with reliable bar classifications. The same classifications are also used to assign a bar type (no bar, weak bar or strong bar) to every galaxy. The galaxy had no bar if $p_{\textrm{strong bar + weak bar}} < 0.5$. If this was not the case and if $p_{\textrm{weak bar}} \geq p_{\textrm{strong bar}}$, then the galaxy had a weak bar. Otherwise, it had a strong bar. This classification scheme was used before in \citet{geron_2021} and is shown in Table \ref{tab:bar_conditions}. The galaxies that are identified as unbarred were removed from our sample, which reduced the sample size to \todo{1,679} barred galaxies. 

\begin{table}
    \centering
    \caption{The vote fractions of the galaxy are used to determine its bar type (no bar, weak bar or strong bar), according to the following scheme. This method of classification is identical to the one in \citet{geron_2021}.}
    \label{tab:bar_conditions}
    \begin{tabular}{ccl}
    \hline
    \hline
    Condition 1 & Condition 2 & Result \\
    \hline
    $p_{\textrm{strong bar + weak bar}} < 0.5$ & N/A & No bar\\
    $p_{\textrm{strong bar + weak bar}} \geq 0.5$ & $p_{\textrm{strong bar}} < p_{\textrm{weak bar}}$ & Weak bar\\
    $p_{\textrm{strong bar + weak bar}} \geq 0.5$ & $p_{\textrm{strong bar}} \geq p_{\textrm{weak bar}}$ & Strong bar\\
    \hline
    \end{tabular}
\end{table}

In order to avoid selection effects, we work with a volume-limited sample by imposing additional thresholds on the redshift ($0.01 < z < 0.05$) and absolute r-band magnitude ($M_{\textrm{r}} <-18.96$, values obtained from the NASA-Sloan Atlas), which removed \todo{519} galaxies from our sample. Limitations of the TW method (see Section \ref{sec:tw_limits}) also impose a few additional thresholds on our sample selection. The TW method is also not developed for galaxies with irregular kinematics, as one of the main assumptions of the TW method is the existence of a well-defined pattern speed. The stellar velocity field of every galaxy was inspected by eye and \todo{475} irregular galaxies were removed. Additionally, the bar cannot align with the disc major or minor axis. Thus, galaxies where the PA of the bar was within 10$^{\circ}$ of the major or minor axis of the galaxy were removed from our sample, which affected \todo{193} galaxies. The TW method only works on galaxies with intermediate inclination, so we limit our sample to galaxies with inclinations between 20$^{\circ}$ and 70$^{\circ}$, which removed a further \todo{32} galaxies.

As our methodology requires to reliably perform a linear fit in the $\left<V\right>$ against $\left<X\right>$ plots, we require each galaxy to have at least three pseudo-slits. Similarly, to ensure the robustness of the linear fit, we use normalized root mean squared error (NRMSE) to estimate the fit quality. We only included targets that had a median NRMSE of all the MC iterations lower than \todo{0.2}. Finally, we fit the rotation curve of the galaxy with a two-parameter arctan function (see Section \ref{sec:rotation_curve}). However, some galaxies are not described correctly by this function, especially galaxies with highly irregular kinematics (which have mostly already been removed by this point). Thus, a threshold of median NRMSE < \todo{0.2} is imposed on this fit as well. These threshold values for NRMSE were chosen after careful visual inspection of their fits. Additionally, targets where more than \todo{10}\% of the MC iterations were unable to provide a value for pattern speed (e.g., due to not being able to place enough pseudo-slits), were excluded as well. Applying these last thresholds result in a final sample that contains \todo{225} galaxies, with \todo{122} strongly barred and \todo{103} weakly barred galaxies. 

Reliable bar pattern speed estimates were obtained for all these targets. However, it was found that for a small subset of these targets, especially for those with low pattern speeds, estimating the corotation radius and $\mathcal{R}$ is difficult. This was because the corotation radius was so high, that it fell far outside the MaNGA field of view. It was judged that extrapolating the velocity curves too much results in unreliable estimates for the corotation radius. Therefore, we excluded any targets where we had to extrapolate by more than a factor of two. This affected \todo{15} of our \todo{225} galaxies.


\section{Results}
\label{sec:results}

\subsection {Bar pattern speeds, corotation radii and $\mathcal{R}$}

\begin{figure*}
	\includegraphics[width=\textwidth]{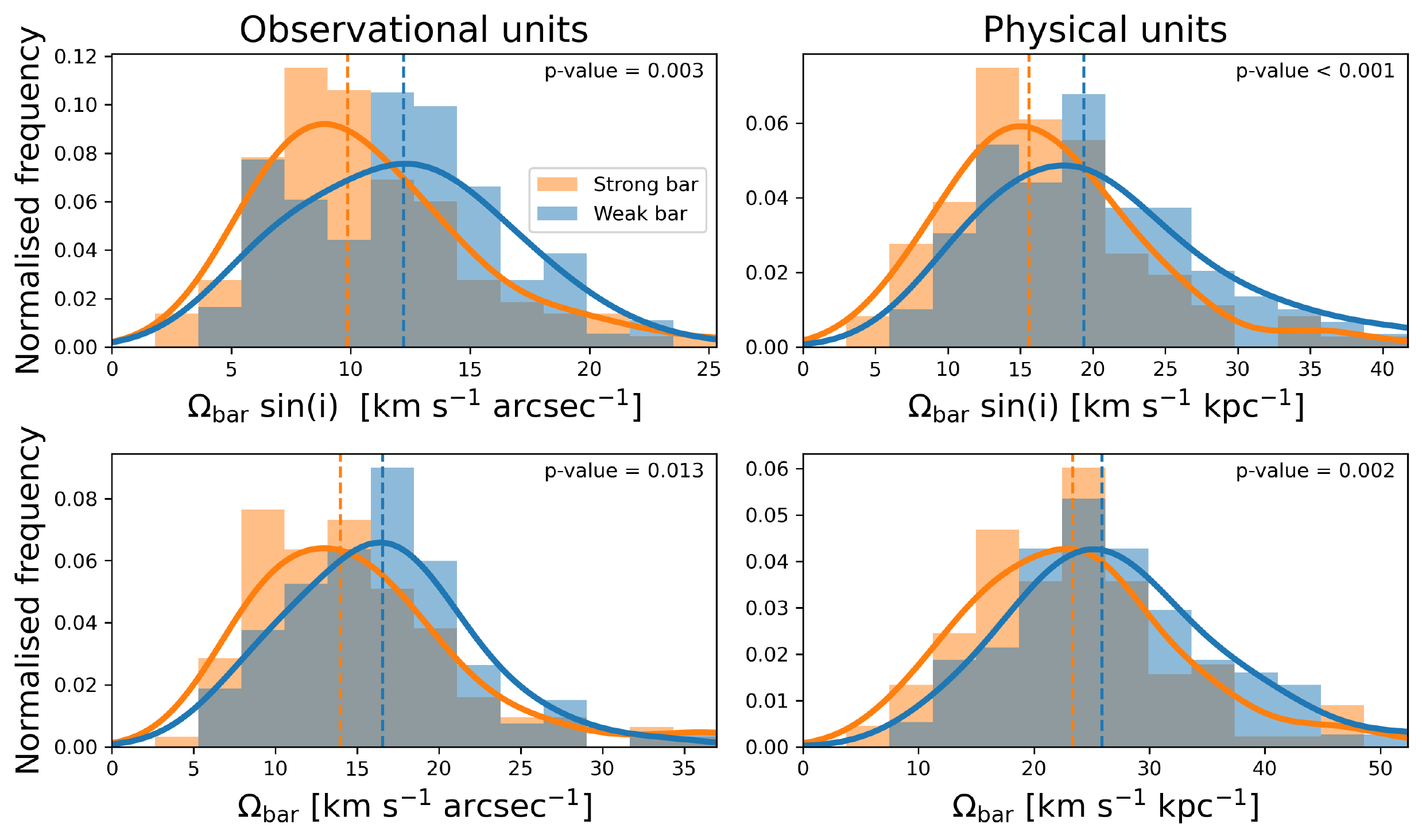}
    \caption{The final median values for $\Omega_{\rm b} \sin \left( i \right)$ (top row) and $\Omega_{\rm b}$ (bottom row) for every galaxy, after doing a Monte Carlo simulation of 1,000 iterations. The sample is divided into strongly barred (orange) and weakly barred (blue). The MC is done over observational units (which include arcsec), which are afterwards converted to kpc. The left column shows the results for the observational units, while the right column shows the results for the physical units. The vertical dashed lines show the median values for every histogram. The full lines are kernel density estimates of these histograms, using a Gaussian kernel. The p-value of a two-sample Anderson-Darling test is shown inside each subplot, with the null hypothesis being that the two samples are drawn from the same population. We see that, on average, strongly barred galaxies have significantly lower bar pattern speeds, despite there being significant overlap between the two populations. The p-value of the comparisons in physical units is $<$0.001-0.002 (which corresponds to 3.1-3.3$\sigma$).}
    \label{fig:omega_hists}
\end{figure*}

The final bar pattern speeds of all our weakly and strongly barred galaxies are shown in Figure \ref{fig:omega_hists}. As mentioned in Section \ref{sec:inclination}, inclination measurements are prone to biases, so to be cautious we show both $\Omega_{\rm b} \sin \left( i \right)$ (top row) and $\Omega_{\rm b}$ (bottom row). Additionally, the pattern speed is measured in observational units (km s$^{-1}$ arcsec$^{-1}$, left column), which are then converted to physical units (km s$^{-1}$ kpc$^{-1}$, right column).

Despite the distributions for the weakly and strongly barred samples overlapping considerably, an Anderson-Darling test reveals that they are still significantly different, with strongly barred galaxies having lower average pattern speeds than weak barred galaxies. The p-values for the $\Omega_{\rm b} \sin \left( i \right)$ distributions are \todo{0.003} \todo{(which corresponds to 3.0$\sigma$)} and \todo{$<$0.001} \todo{($>$3.3$\sigma$)} for the plots using observational units and physical units, respectively. The p-values for the $\Omega_{\rm b}$ plots are \todo{0.013} \todo{(2.5$\sigma$)} and \todo{0.002} \todo{(3.1$\sigma$)}, respectively.

We can conclude that strongly barred galaxies have significantly lower bar pattern speeds than weakly barred galaxies. The median, together with the 16$^{\rm th}$ and 84$^{\rm th}$ percentiles, is $\Omega_{\rm b} = \todo{23.36^{+9.25}_{-8.1}}$ km s$^{-1}$ kpc$^{-1}$ for strongly barred galaxies and $\Omega_{\rm b} = \todo{25.91^{+10.42}_{-7.26}}$ km s$^{-1}$ kpc$^{-1}$ for weakly barred galaxies.

\begin{figure*}
	\includegraphics[width=\textwidth]{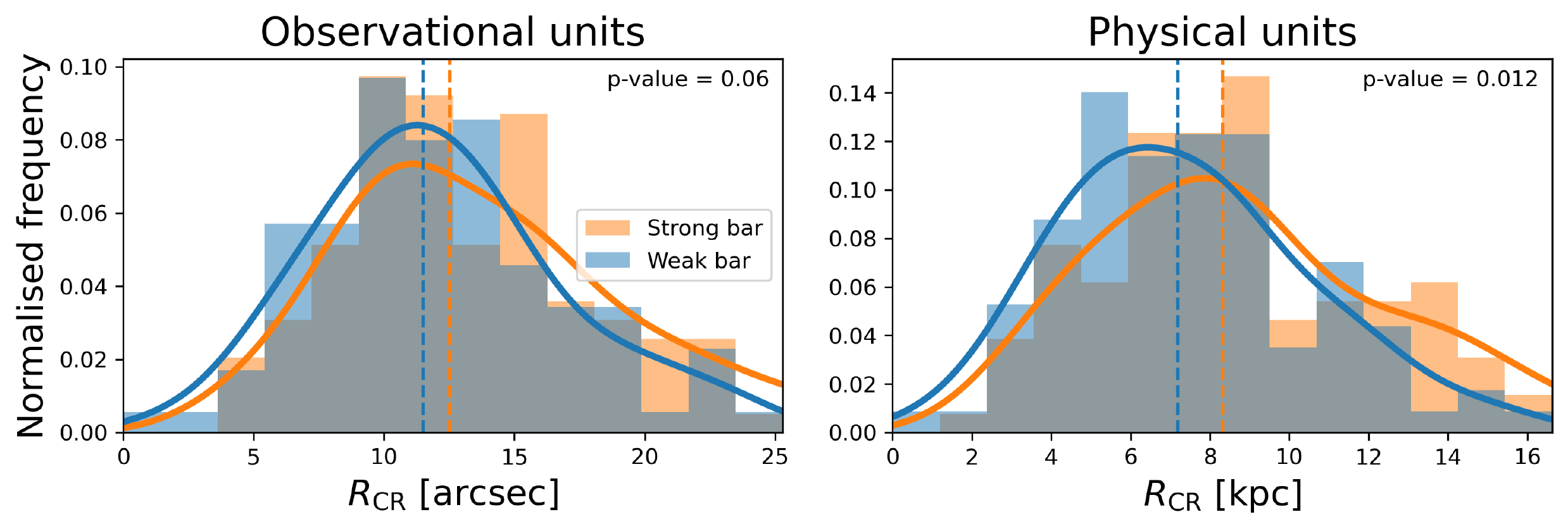}
    \caption{The final median values for R$_{\rm CR}$  for every galaxy, after doing a Monte Carlo simulation of 1,000 iterations. The sample is divided into strongly barred (orange) and weakly barred (blue). The MC is done over observational units (which is in arcsec), which are afterwards converted to kpc. The left column shows the results for the observational units, while the right column shows the results for the physical units. The vertical dashed lines show the median values for every histogram. The full lines are kernel density estimates of these histograms, using a Gaussian kernel. The p-value of a two-sample Anderson-Darling test is shown inside each subplot, with the null hypothesis being that the two samples are drawn from the same population. As the p-value of the comparison in physical units is 0.012 (which corresponds to 2.5$\sigma$), we conclude that we see no significant difference between weak and strong bars in terms of their corotation radii.}
    \label{fig:rcr_hists}
\end{figure*}

The final corotation radii for our target galaxies are shown in Figure \ref{fig:rcr_hists}. The median values are R$_{\rm CR} = \todo{8.33^{+4.57}_{-3.31}}$ kpc and R$_{\rm CR} = \todo{7.19^{+3.82}_{-2.96}}$ kpc for the strongly barred and weakly barred sample, respectively. The results from a two-sample Anderson-Darling test show that the distributions of the corotation radii in physical units are not significantly different between weak and strong bars (p-value = \todo{0.012}; \todo{2.5}$\sigma$), as the significance is below 3$\sigma$.

\begin{figure}
	\includegraphics[width=\columnwidth]{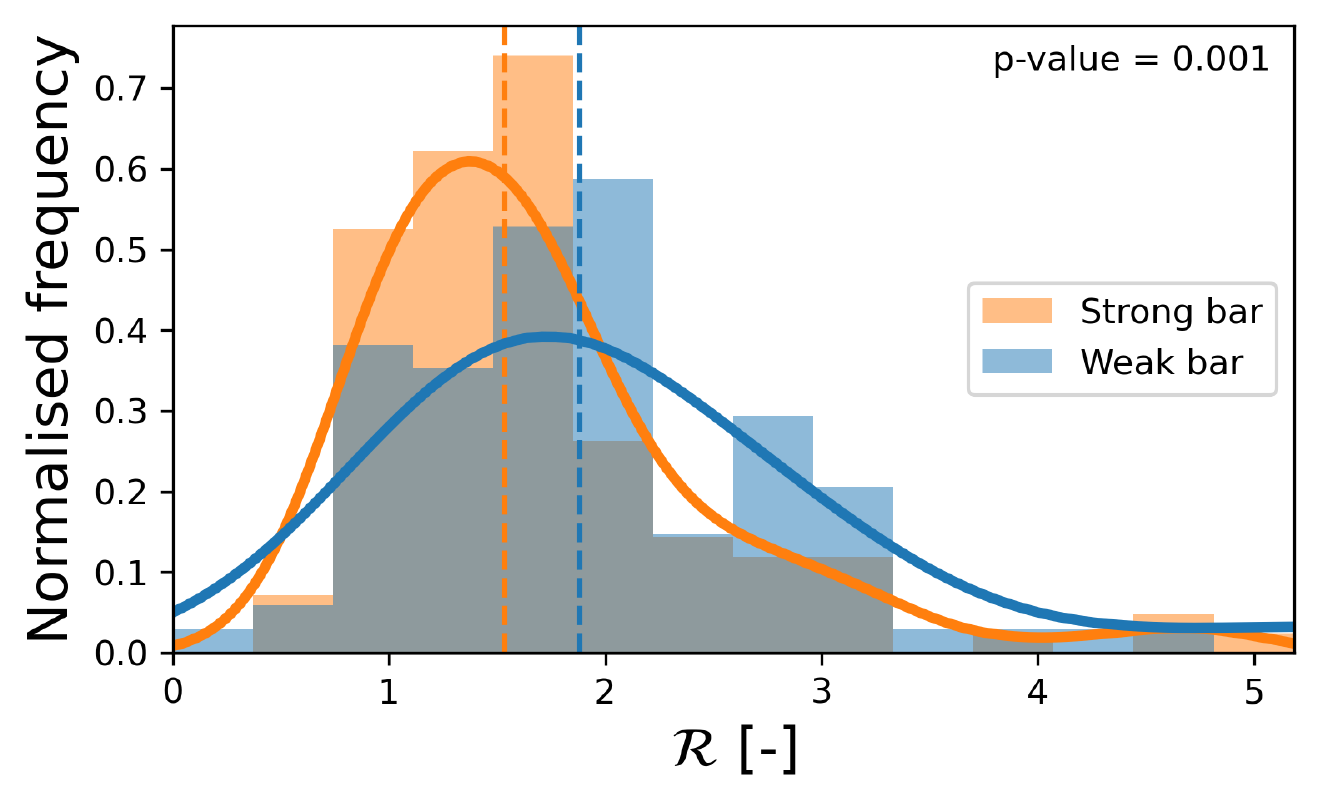}
    \caption{The final median values for $\mathcal{R}$ for every galaxy, after doing a Monte Carlo simulation of 1,000 iterations. The sample is divided into strongly barred (orange) and weakly barred (blue). The vertical dashed lines show the median values for every histogram. The full lines are kernel density estimates of these histograms, using a Gaussian kernel. The p-value of a two-sample Anderson-Darling test is shown inside each subplot, with the null hypothesis being that the two samples are drawn from the same population. We see that strong bars have significantly lower values of $\mathcal{R}$ than weak bars, despite the big overlap (p-value = 0.001; 3.3$\sigma$).}
    \label{fig:r_hist}
\end{figure}

With a p-value of \todo{0.001}, which corresponds to \todo{3.3}$\sigma$, we conclude that strong bars have significantly lower values for $\mathcal{R}$ than weak bars, as shown in Figure \ref{fig:r_hist}. However, please note that the distributions still overlap significantly. The median value for strongly barred galaxies is $\mathcal{R} = \todo{1.53^{+0.74}_{-0.53}}$ and $\mathcal{R} = \todo{1.88^{+1.08}_{-0.75}}$ for weakly barred galaxies.

As mentioned above, $\mathcal{R}$ is used to divide bars into ultrafast ($\mathcal{R} < 1.0$), fast ($1.0 < \mathcal{R} < 1.4$) and slow ($\mathcal{R} > 1.4$). Most bars in our sample seem to be slow bars (\todo{62}\% of our sample), as shown in Table \ref{tab:ultrafast_bars}. This fraction is higher than what most other studies find. Conversely, we find less ultrafast and fast bars (\todo{11}\% and \todo{27}\%, respectively) than most other studies. 

The final values for pattern speeds, corotation radii and $\mathcal{R}$ of 50 randomly selected galaxies is shown in Table \ref{tab:results}. The full table can be found online \href{https://doi.org/10.5281/zenodo.7567945}{here}\footnote{\url{https://doi.org/10.5281/zenodo.7567945}}.

\subsection {Relationship between the parameters}

\begin{figure}
	\includegraphics[width=\columnwidth]{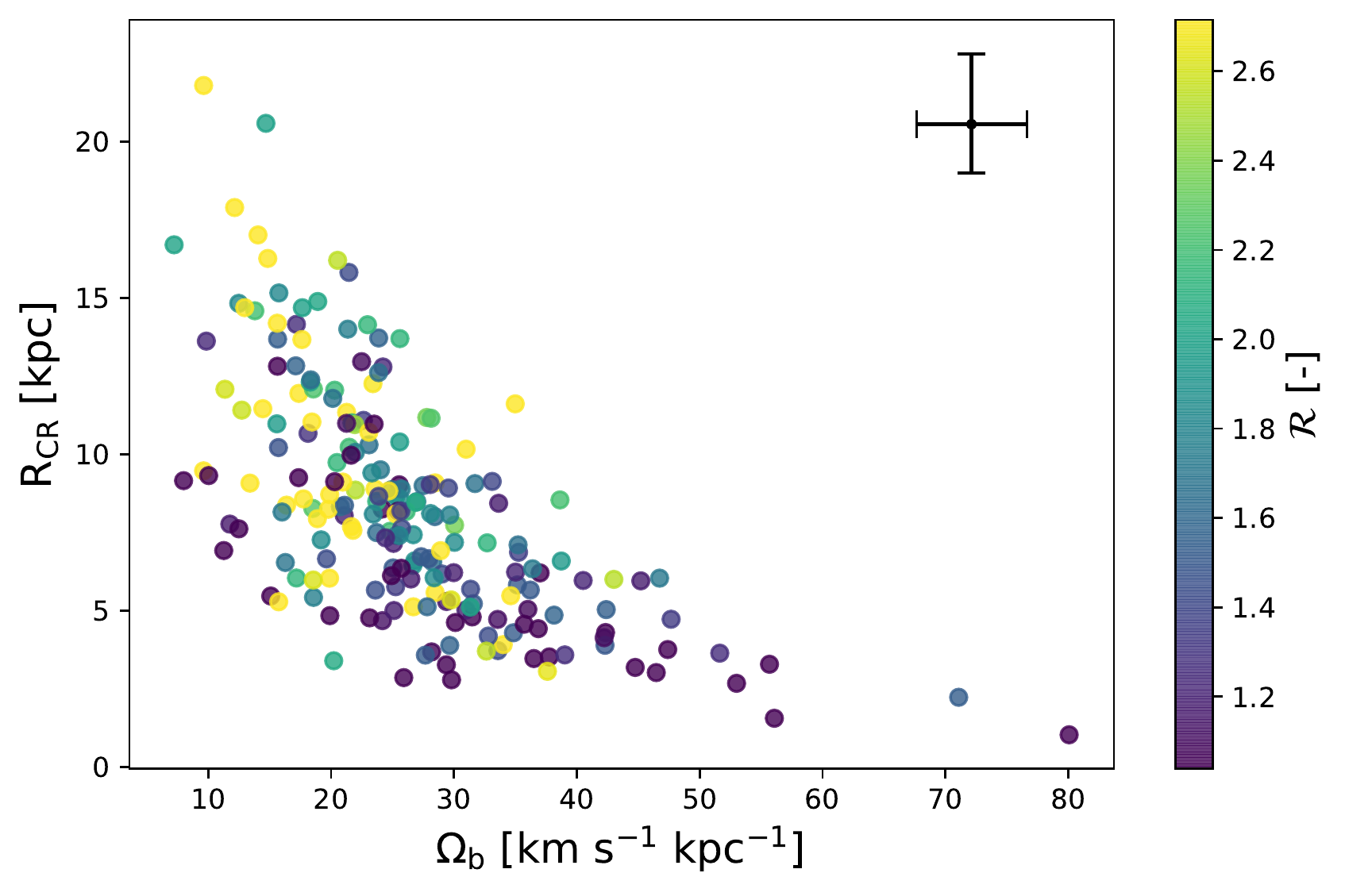}
    \caption{The bar pattern speed ($\Omega_{\rm b}$) of all our targets plotted against the corotation radius (R$_{\rm CR}$). The colour of the data points is determined by $\mathcal{R}$. The median error on the x and y axis is shown in the top-right corner. We can see that $\Omega_{\rm b}$ and R$_{\rm CR}$ are clearly inversely proportional, as expected. Additionally, we see that low $\mathcal{R}$ values cluster at lower values for R$_{\rm CR}$. To aid visualisation, the colours used to indicate $\mathcal{R}$ were capped at the 16$^{\rm th}$ and 84$^{\rm th}$ percentile.}
    \label{fig:Om_Rcr_R}
\end{figure}

The bar pattern speed and corotation radius should be inversely proportional to each other. A higher pattern speed will result in a steeper gradient of the straight line shown in Figure \ref{fig:example_velcurve}, resulting in it intersecting with the rotation curve at a shorter distance, which produces a lower corotation radius (see Section \ref{sec:calc_rcr_r} for more details). This is shown explicitly in Figure \ref{fig:Om_Rcr_R}, where the inversely proportional relationship becomes very clear.

We can also see that galaxies with the highest values of $\mathcal{R}$ tend to have lower values for the bar pattern speed. This also makes sense, as low pattern speeds will result in larger corotation radii, which increases $\mathcal{R}$. Conversely, galaxies with lower $\mathcal{R}$ tend to have lower values for the corotation radius.

\begin{figure}
	\includegraphics[width=\columnwidth]{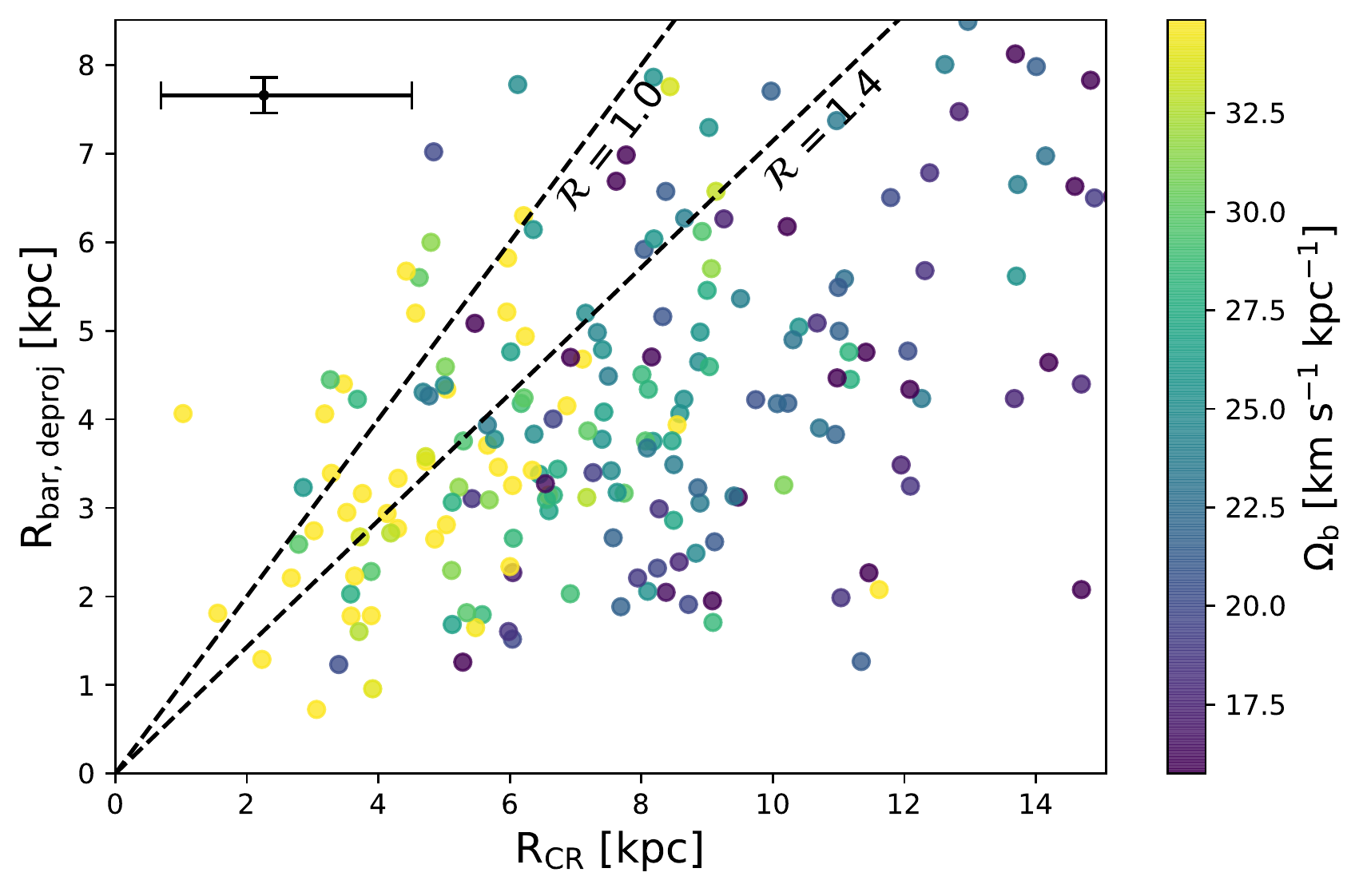}
    \caption{The corotation radius is plotted against the deprojected bar length. As $\mathcal{R} = R_{\rm CR}/R_{\rm bar}$, this figure is divided into three regions: the region with slow bars ($\mathcal{R} > 1.4$), the region with fast bars ($1 < \mathcal{R} < 1.4$) and the region with ultrafast bars ($\mathcal{R} < 1$). The colour indicates the bar pattern speed. The median error on the x and y axis is shown in the top-left corner. To aid visualisation, the colours used to indicate the bar pattern speed were capped at the 16$^{\rm th}$ and 84$^{\rm th}$ percentile.}
    \label{fig:R_plot}
\end{figure}

Figure \ref{fig:R_plot} shows the corotation radius plotted against the bar radius. Lines of equal values of $\mathcal{R}$ are found diagonally over this Figure, as $\mathcal{R}$ is defined as the ratio between the corotation radius and the bar radius. Thus, we can divide this figure into three regions, one with all the ultrafast bars, one with all the fast bars and one with all the slow bars. The bar pattern speed is shown with the colour gradient. The galaxies with the fastest pattern speeds mostly have low values for the bar radius and corotation radius. The galaxies with the lowest pattern speeds tend to have higher values for the corotation radius, as well as higher values for $\mathcal{R}$.

\begin{figure*}
	\includegraphics[width=\textwidth]{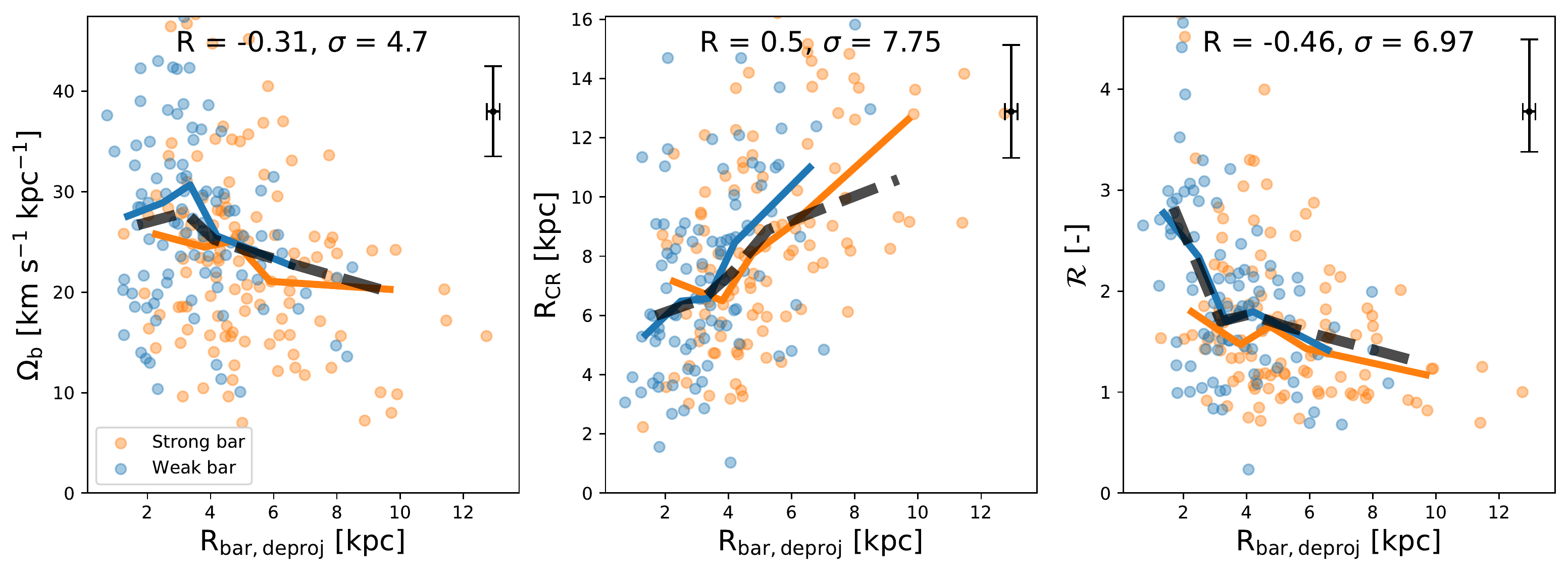}
    \caption{The bar pattern speed (left panel), corotation radius (middle panel) and $\mathcal{R}$ (right panel) against the bar radius. All strongly barred galaxies are coloured orange, while all weakly barred galaxies are coloured blue. The median trend for the weakly and strongly barred galaxies is shown with the blue and orange full lines, respectively. Additionally, the general median trend of all barred galaxies is shown in the dashed black line. The Spearman correlation coefficient, R, and its significance, $\sigma$, are shown in every subplot. The median error on the x and y axis is shown in the top-right corner. We see that the pattern speed and $\mathcal{R}$ decrease with bar length, while the corotation radius increases.}
    \label{fig:against_barlen}
\end{figure*}

The bar radius is plotted against the bar pattern speed, corotation radius and $\mathcal{R}$ in Figure \ref{fig:against_barlen}. We find that the bar pattern speed decreases as bar radius increases. Though the Spearman correlation index is quite small (R = \todo{-0.31}) due to the high amounts of scatter, its significance is high (\todo{4.7$\sigma$}). A more careful look reveals that all the largest bars have lower values for the bar pattern speeds. Conversely, all bars with higher values for their pattern speed are relatively short.

The corotation radius increases with bar radius (R = \todo{0.5}; \todo{7.75}$\sigma$). This is because larger bars need to have a larger corotation radius, as a bar can only grow up to its corotation radius. $\mathcal{R}$ is observed to decrease with bar radius (R = \todo{-0.46}; \todo{6.97}$\sigma$). However, $\mathcal{R}$ is very sensitive to correct bar length estimates, so this trend could merely be a reflection of that. Interestingly, the median trends for the weakly and strongly barred subsamples are very similar to each other.

\begin{figure*}
	\includegraphics[width=\textwidth]{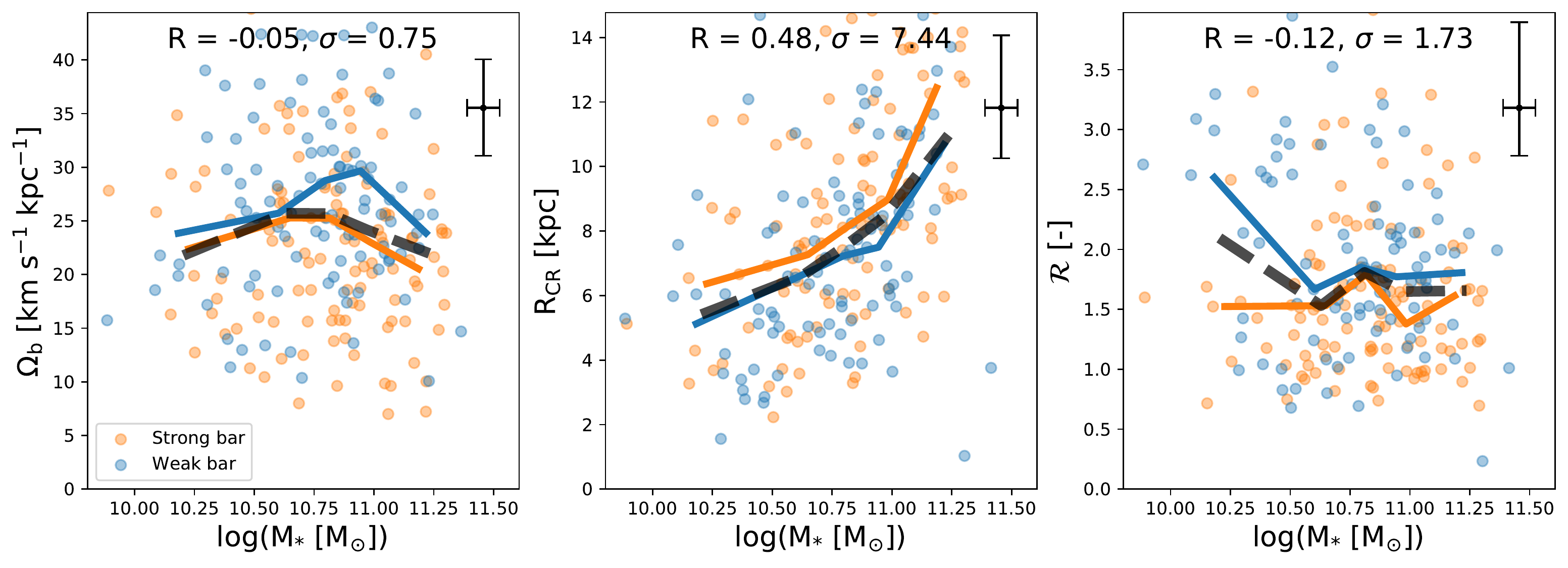}
    \caption{The bar pattern speed (left panel), corotation radius (middle panel) and $\mathcal{R}$ (right panel) against the stellar mass. All strongly barred galaxies are coloured orange, while all weakly barred galaxies are coloured blue. The median trend for the weakly and strongly barred galaxies is shown with the blue and orange full lines, respectively. Additionally, the general median trend of all barred galaxies is shown in the dashed black line. The Spearman correlation coefficient, R, and its significance, $\sigma$, are shown in every subplot. The median error on the x and y axis is shown in the top-right corner. We see that the pattern speed and $\mathcal{R}$ do not show a significant trend with stellar mass, while the corotation radius seems to increase with stellar mass.}
    \label{fig:against_stellarmass}
\end{figure*}

Many properties of galaxies vary with stellar mass \citep{brinchmann_2000,brinchmann_2004,noeske_2007,lara-lopez_2010} and bars, especially stronger bars, are known to appear more often in massive galaxies \citep{masters_2012,cervantessodi_2017,geron_2021}. In Figure \ref{fig:against_stellarmass}, we plot the stellar mass against the bar pattern speed, corotation radius and $\mathcal{R}$ to see if any of these parameters are correlated with stellar mass as well. We see that the pattern speed and $\mathcal{R}$ do not correlate with stellar mass (\todo{R = -0.05; 0.75$\sigma$ and R = -0.12; 1.73$\sigma$}, respectively). This shows that the differences we observed in pattern speed in Figures \ref{fig:omega_hists} and \ref{fig:r_hist} are not due to differences in stellar mass of our targets. Interestingly, the corotation radius does increase with stellar mass (\todo{R = 0.48; 7.44$\sigma$}). This is because more massive galaxies tend to host stronger and longer bars, which tend to have larger corotation radii, as shown in the middle panel of Figure \ref{fig:against_barlen}.

\begin{table*}
    \centering
    \captionsetup{justification=centerlast}
    \caption{The bar pattern speeds, corotation radii and $\mathcal{R}$ for 50 randomly selected bars. The full table can be found online \href{https://doi.org/10.5281/zenodo.7567945}{here}.}

    \begin{tabular}{llllll}
        \hline
        \hline
        Plate-ifu & $\Omega_{\rm b}$ [km s$^{-1}$ arcsec$^{-1}$] & $\Omega_{\rm b}$ [km s$^{-1}$ kpc$^{-1}$] & R$_{\rm CR}$ [arcsec] & R$_{\rm CR}$ [kpc] & $\mathcal{R}$ [-]\\
        \hline
        11014-12705 & 9.40$_{-1.19}^{+0.99}$ & 27.82$_{-3.52}^{+2.92}$ & 15.16$_{-2.06}^{+2.89}$ & 5.12$_{-0.70}^{+0.98}$ & 1.60$_{-0.29}^{+0.50}$ \\[0.09cm]
        9504-3704 & 34.27$_{-10.60}^{+14.39}$ & 36.50$_{-11.29}^{+15.32}$ & 3.70$_{-1.14}^{+2.25}$ & 3.47$_{-1.07}^{+2.11}$ & 0.85$_{-0.28}^{+0.57}$ \\[0.09cm]
        8245-12702 & 15.36$_{-1.68}^{+0.85}$ & 22.48$_{-2.46}^{+1.24}$ & 18.98$_{-1.51}^{+3.15}$ & 12.97$_{-1.03}^{+2.16}$ & 1.09$_{-0.21}^{+0.74}$ \\[0.09cm]
        11979-12703 & 14.50$_{-4.37}^{+4.41}$ & 20.48$_{-6.17}^{+6.23}$ & 13.75$_{-3.83}^{+6.81}$ & 9.74$_{-2.71}^{+4.82}$ & 2.14$_{-0.65}^{+1.06}$ \\[0.09cm]
        9027-12704 & 5.51$_{-0.90}^{+2.89}$ & 7.99$_{-1.30}^{+4.19}$ & 13.27$_{-6.68}^{+3.56}$ & 9.16$_{-4.61}^{+2.46}$ & 0.82$_{-0.30}^{+0.26}$ \\[0.09cm]
        8079-9101 & 13.18$_{-1.92}^{+1.93}$ & 28.09$_{-4.09}^{+4.10}$ & 17.28$_{-3.10}^{+4.17}$ & 8.11$_{-1.45}^{+1.96}$ & 1.84$_{-0.43}^{+0.65}$ \\[0.09cm]
        8624-9102 & 12.50$_{-6.80}^{+4.01}$ & 21.98$_{-11.95}^{+7.05}$ & 15.57$_{-4.61}^{+21.07}$ & 8.86$_{-2.62}^{+11.99}$ & 2.53$_{-0.86}^{+5.06}$ \\[0.09cm]
        10220-9101 & 20.55$_{-3.36}^{+2.06}$ & 28.44$_{-4.65}^{+2.86}$ & 11.09$_{-1.50}^{+3.11}$ & 8.01$_{-1.08}^{+2.25}$ & 1.60$_{-0.31}^{+0.69}$ \\[0.09cm]
        11979-9101 & 20.83$_{-5.28}^{+4.32}$ & 23.87$_{-6.05}^{+4.95}$ & 9.92$_{-2.24}^{+4.36}$ & 8.66$_{-1.95}^{+3.80}$ & 1.37$_{-0.36}^{+0.65}$ \\[0.09cm]
        8723-12701 & 19.70$_{-2.78}^{+2.80}$ & 25.61$_{-3.62}^{+3.65}$ & 11.16$_{-1.78}^{+2.30}$ & 8.59$_{-1.37}^{+1.77}$ & 1.85$_{-0.37}^{+0.55}$ \\[0.09cm]
        9881-12704 & 6.74$_{-1.09}^{+2.51}$ & 12.49$_{-2.02}^{+4.64}$ & 14.11$_{-5.72}^{+4.08}$ & 7.62$_{-3.09}^{+2.20}$ & 1.00$_{-0.31}^{+0.32}$ \\[0.09cm]
        11868-12703 & 8.45$_{-4.42}^{+3.67}$ & 9.84$_{-5.15}^{+4.27}$ & 15.87$_{-5.86}^{+20.56}$ & 13.62$_{-5.03}^{+17.65}$ & 1.23$_{-0.67}^{+1.60}$ \\[0.09cm]
        11965-9102 & 16.30$_{-0.72}^{+0.66}$ & 25.25$_{-1.11}^{+1.03}$ & 8.93$_{-0.58}^{+0.69}$ & 5.77$_{-0.37}^{+0.45}$ & 1.34$_{-0.23}^{+0.35}$ \\[0.09cm]
        10492-12702 & 18.34$_{-0.96}^{+1.01}$ & 33.58$_{-1.75}^{+1.85}$ & 6.82$_{-0.39}^{+0.36}$ & 3.72$_{-0.22}^{+0.20}$ & 1.43$_{-0.10}^{+0.12}$ \\[0.09cm]
        8257-3703 & 23.40$_{-2.10}^{+1.89}$ & 46.47$_{-4.16}^{+3.76}$ & 6.00$_{-0.75}^{+0.95}$ & 3.02$_{-0.38}^{+0.48}$ & 0.92$_{-0.14}^{+0.17}$ \\[0.09cm]
        8602-12705 & 7.94$_{-3.42}^{+1.43}$ & 12.48$_{-5.37}^{+2.25}$ & 23.31$_{-4.08}^{+20.25}$ & 14.83$_{-2.59}^{+12.88}$ & 1.83$_{-0.42}^{+2.14}$ \\[0.09cm]
        8323-12705 & 14.47$_{-2.55}^{+1.95}$ & 21.45$_{-3.78}^{+2.89}$ & 23.46$_{-4.95}^{+8.73}$ & 15.82$_{-3.34}^{+5.89}$ & 1.42$_{-0.52}^{+1.18}$ \\[0.09cm]
        8442-9102 & 13.10$_{-3.15}^{+4.56}$ & 28.18$_{-6.77}^{+9.80}$ & 7.92$_{-1.90}^{+2.16}$ & 3.68$_{-0.88}^{+1.00}$ & 1.06$_{-0.28}^{+0.33}$ \\[0.09cm]
        10492-6102 & 23.05$_{-4.40}^{+4.48}$ & 38.74$_{-7.40}^{+7.54}$ & 11.07$_{-2.34}^{+3.26}$ & 6.59$_{-1.39}^{+1.94}$ & 1.94$_{-0.54}^{+1.15}$ \\[0.09cm]
        11956-12702 & 12.37$_{-0.46}^{+0.42}$ & 15.63$_{-0.59}^{+0.53}$ & 16.20$_{-0.71}^{+0.80}$ & 12.82$_{-0.56}^{+0.63}$ & 1.00$_{-0.11}^{+0.11}$ \\[0.09cm]
        10226-3704 & 28.26$_{-8.25}^{+7.75}$ & 35.22$_{-10.29}^{+9.66}$ & 8.85$_{-3.15}^{+5.36}$ & 7.11$_{-2.53}^{+4.30}$ & 1.66$_{-0.63}^{+1.00}$ \\[0.09cm]
        8145-3704 & 25.85$_{-2.46}^{+3.06}$ & 46.74$_{-4.45}^{+5.53}$ & 10.92$_{-1.97}^{+1.84}$ & 6.04$_{-1.09}^{+1.02}$ & 1.71$_{-0.29}^{+0.26}$ \\[0.09cm]
        8456-6101 & 16.73$_{-2.92}^{+3.04}$ & 35.72$_{-6.23}^{+6.50}$ & 9.75$_{-3.34}^{+4.43}$ & 4.57$_{-1.56}^{+2.08}$ & 0.97$_{-0.31}^{+0.44}$ \\[0.09cm]
        12651-9101 & 17.99$_{-1.32}^{+7.31}$ & 20.29$_{-1.49}^{+8.24}$ & 10.29$_{-3.60}^{+0.99}$ & 9.13$_{-3.20}^{+0.88}$ & 0.70$_{-0.15}^{+0.12}$ \\[0.09cm]
        8324-9101 & 12.49$_{-2.46}^{+3.31}$ & 21.65$_{-4.26}^{+5.74}$ & 13.33$_{-3.90}^{+4.29}$ & 7.69$_{-2.25}^{+2.47}$ & 3.52$_{-1.29}^{+2.08}$ \\[0.09cm]
        8084-6101 & 14.46$_{-1.51}^{+1.52}$ & 34.84$_{-3.65}^{+3.65}$ & 10.34$_{-1.39}^{+1.67}$ & 4.29$_{-0.58}^{+0.69}$ & 1.52$_{-0.27}^{+0.40}$ \\[0.09cm]
        11004-12704 & 11.63$_{-3.80}^{+3.90}$ & 23.60$_{-7.71}^{+7.91}$ & 11.49$_{-3.46}^{+7.17}$ & 5.66$_{-1.71}^{+3.53}$ & 1.45$_{-0.45}^{+0.93}$ \\[0.09cm]
        9886-12701 & 8.76$_{-3.85}^{+3.02}$ & 18.88$_{-8.29}^{+6.51}$ & 17.12$_{-5.44}^{+16.59}$ & 7.94$_{-2.52}^{+7.70}$ & 3.07$_{-1.07}^{+4.55}$ \\[0.09cm]
        8602-12701 & 15.91$_{-2.48}^{+4.39}$ & 29.55$_{-4.61}^{+8.15}$ & 16.57$_{-2.72}^{+2.54}$ & 8.92$_{-1.46}^{+1.37}$ & 1.40$_{-0.29}^{+0.33}$ \\[0.09cm]
        9190-12703 & 15.70$_{-4.50}^{+1.66}$ & 35.16$_{-10.08}^{+3.72}$ & 13.04$_{-1.80}^{+7.20}$ & 5.82$_{-0.80}^{+3.22}$ & 1.51$_{-0.34}^{+1.02}$ \\[0.09cm]
        8978-9101 & 19.17$_{-6.35}^{+3.04}$ & 31.58$_{-10.45}^{+5.01}$ & 8.61$_{-1.80}^{+6.29}$ & 5.23$_{-1.09}^{+3.82}$ & 1.53$_{-0.35}^{+1.17}$ \\[0.09cm]
        8137-9102 & 15.21$_{-3.58}^{+2.08}$ & 24.43$_{-5.75}^{+3.35}$ & 11.77$_{-4.07}^{+8.36}$ & 7.33$_{-2.53}^{+5.21}$ & 1.24$_{-0.44}^{+1.23}$ \\[0.09cm]
        8596-12702 & 36.23$_{-2.97}^{+10.54}$ & 47.67$_{-3.91}^{+13.88}$ & 6.22$_{-1.80}^{+0.70}$ & 4.73$_{-1.37}^{+0.53}$ & 1.34$_{-0.16}^{+0.16}$ \\[0.09cm]
        12495-6102 & 12.70$_{-5.25}^{+2.91}$ & 23.57$_{-9.75}^{+5.41}$ & 16.51$_{-4.36}^{+13.57}$ & 8.89$_{-2.35}^{+7.31}$ & 2.87$_{-1.39}^{+1.98}$ \\[0.09cm]
        9028-12702 & 15.20$_{-1.49}^{+1.52}$ & 25.25$_{-2.47}^{+2.53}$ & 14.37$_{-1.64}^{+2.04}$ & 8.65$_{-0.99}^{+1.23}$ & 1.89$_{-0.22}^{+0.25}$ \\[0.09cm]
        8619-12701 & 19.39$_{-1.84}^{+1.86}$ & 33.12$_{-3.14}^{+3.18}$ & 15.60$_{-3.07}^{+2.90}$ & 9.13$_{-1.80}^{+1.70}$ & 1.40$_{-0.29}^{+0.28}$ \\[0.09cm]
        11834-12705 & 12.41$_{-3.37}^{+1.98}$ & 14.84$_{-4.03}^{+2.37}$ & 19.45$_{-3.17}^{+8.34}$ & 16.27$_{-2.65}^{+6.97}$ & 2.77$_{-0.71}^{+1.44}$ \\[0.09cm]
        8982-6104 & 19.93$_{-2.62}^{+3.59}$ & 28.38$_{-3.73}^{+5.12}$ & 8.62$_{-1.24}^{+1.57}$ & 6.05$_{-0.87}^{+1.10}$ & 1.85$_{-0.59}^{+0.91}$ \\[0.09cm]
        7962-12704 & 8.96$_{-3.26}^{+3.44}$ & 19.87$_{-7.23}^{+7.62}$ & 13.40$_{-4.36}^{+9.69}$ & 6.04$_{-1.97}^{+4.37}$ & 2.99$_{-1.11}^{+3.16}$ \\[0.09cm]
        9095-9102 & 20.19$_{-7.82}^{+11.35}$ & 31.33$_{-12.13}^{+17.62}$ & 7.94$_{-3.44}^{+5.99}$ & 5.11$_{-2.22}^{+3.86}$ & 2.01$_{-0.87}^{+2.48}$ \\[0.09cm]
        8947-9101 & 8.89$_{-3.00}^{+3.27}$ & 9.62$_{-3.24}^{+3.54}$ & 10.25$_{-3.25}^{+5.33}$ & 9.47$_{-3.01}^{+4.93}$ & 2.83$_{-1.04}^{+2.15}$ \\[0.09cm]
        9089-12704 & 14.92$_{-1.70}^{+1.11}$ & 23.50$_{-2.67}^{+1.75}$ & 17.28$_{-1.71}^{+3.13}$ & 10.97$_{-1.09}^{+1.99}$ & 0.97$_{-0.27}^{+0.53}$ \\[0.09cm]
        11976-12705 & 8.26$_{-3.70}^{+4.13}$ & 13.39$_{-6.00}^{+6.70}$ & 14.72$_{-6.03}^{+13.98}$ & 9.08$_{-3.72}^{+8.62}$ & 4.42$_{-1.56}^{+4.75}$ \\[0.09cm]
        11743-6104 & 15.96$_{-3.79}^{+2.70}$ & 27.33$_{-6.49}^{+4.62}$ & 11.52$_{-3.08}^{+5.70}$ & 6.73$_{-1.80}^{+3.33}$ & 1.51$_{-0.39}^{+0.76}$ \\[0.09cm]
        10492-9101 & 9.97$_{-0.76}^{+1.61}$ & 18.51$_{-1.41}^{+2.99}$ & 15.36$_{-2.53}^{+1.25}$ & 8.27$_{-1.36}^{+0.67}$ & 2.26$_{-0.27}^{+1.01}$ \\[0.09cm]
        8244-3702 & 9.06$_{-3.49}^{+3.29}$ & 16.38$_{-6.31}^{+5.95}$ & 15.14$_{-4.75}^{+10.82}$ & 8.38$_{-2.63}^{+5.98}$ & 4.52$_{-1.65}^{+3.71}$ \\[0.09cm]
        11834-6103 & 12.23$_{-5.44}^{+5.24}$ & 14.44$_{-6.42}^{+6.18}$ & 13.53$_{-4.59}^{+11.99}$ & 11.46$_{-3.89}^{+10.16}$ & 4.82$_{-1.94}^{+4.96}$ \\[0.09cm]
        8621-12704 & 17.66$_{-3.79}^{+2.08}$ & 21.36$_{-4.59}^{+2.52}$ & 16.94$_{-2.62}^{+6.14}$ & 14.01$_{-2.17}^{+5.07}$ & 1.76$_{-0.37}^{+0.88}$ \\[0.09cm]
        8615-6104 & 17.68$_{-3.32}^{+2.61}$ & 25.60$_{-4.81}^{+3.78}$ & 19.84$_{-3.48}^{+5.83}$ & 13.70$_{-2.40}^{+4.03}$ & 2.13$_{-0.50}^{+1.11}$ \\[0.09cm]
        12622-9102 & 22.67$_{-19.23}^{+12.44}$ & 26.98$_{-22.88}^{+14.80}$ & 10.10$_{-4.89}^{+73.55}$ & 8.49$_{-4.11}^{+61.81}$ & 1.74$_{-0.85}^{+22.56}$ \\[0.09cm]
        and \todo{175} more rows...\\
        \hline
    \end{tabular}

    \label{tab:results}
\end{table*}

\subsection{Quenching}

\begin{figure*}
	\includegraphics[width=\textwidth]{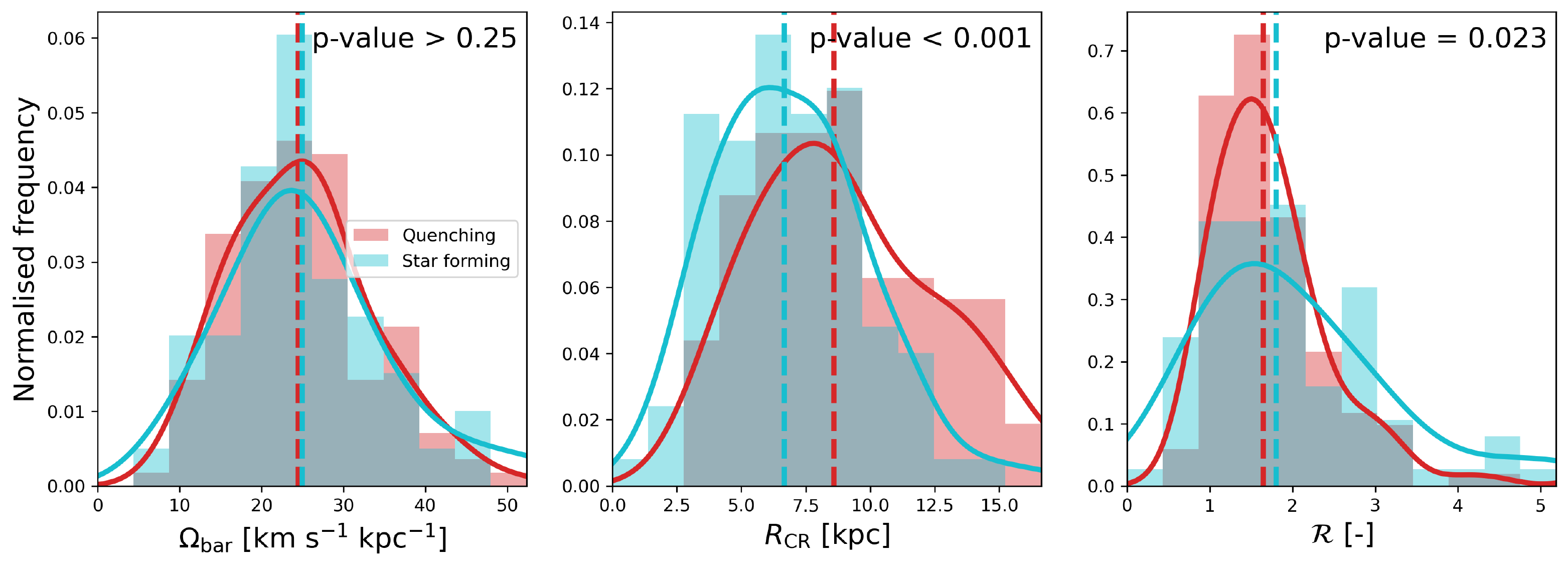}
    \caption{The final median values for bar pattern speed (left panel), corotation radius (middle panel) and $\mathcal{R}$ (right panel) after doing a Monte Carlo simulation of 1,000 iterations. The sample is divided into quenching galaxies (red) and star forming galaxies (blue). The vertical dashed lines show the median values for every histogram, while the full lines are kernel density estimates of these histograms, using a Gaussian kernel. The p-value of a two-sample Anderson-Darling test is shown in the top-right corner of every subplot. The null hypothesis is that the two samples in each subplot are drawn from the same population. We can see that the quenching and star forming subsamples are not significantly different in terms of pattern speed and $\mathcal{R}$, but are in terms of corotation radius.}
    \label{fig:deltams_hists}
\end{figure*}

Our previous results reveal a complicated interplay between bar pattern speed, corotation radii, $\mathcal{R}$, bar length and bar type. It is also known that strong bars are more often found in red sequence galaxies \citep{masters_2012,vera_2016,geron_2021}. This suggest a potential link between these dynamical parameters and quenching. Galaxies can be classified as star forming or quiescent based on their location on the SFR-stellar mass plane. We can use the star formation main sequence (SFMS) defined in \citet{belfiore_2018}:

\begin{equation}
    \log{ \left( \textrm{SFR}/ \textrm{M}_{\odot}\,\textrm{yr}^{-1} \right) } = (0.73\,\pm\,0.03) \log{ \left(\textrm{M}_{\textrm{*}}/\textrm{M}_{\odot} \right)} - (7.33\,\pm\,0.29) \;,
	\label{eq:sfr_mass_split}
\end{equation}

and assume that all galaxies that are 1$\sigma$ ($=0.39$ dex) below this line are undergoing quenching and everything else is star forming \citep{belfiore_2018}. \todo{57}\% of our barred galaxies are quenching, whereas \todo{43}\% are star forming. The bar pattern speeds, corotation radii and values for $\mathcal{R}$ for all the star forming and quenching galaxy subsamples are shown in Figure \ref{fig:deltams_hists}. An Anderson-Darling test between the subsamples shows that the pattern speed and $\mathcal{R}$ are not significantly different (both are \todo{$<$3$\sigma$}). However, with a p-value of $<$\todo{0.001}, which corresponds to \todo{$>$3.3$\sigma$}, the corotation radii are significantly different. Thus, our results suggest that quenching galaxies tend to have significantly higher corotation radii than star forming galaxies.

\subsection{Comparison with other work}
\label{sec:comparison_other_work}

Various other studies have also tried to measure bar pattern speeds, corotation radii and $\mathcal{R}$. \citet{rautiainen_2008} determine pattern speeds, corotation radii and $\mathcal{R}$ for a sample of 38 galaxies with data from the Ohio State University Bright Spiral Galaxy Survey \citep{eskridge_2002}. \citet{aguerri_2015} used the Calar Alto Legacy Integral Field Area (CALIFA, \citealt{sanchez_2012}) survey on 15 galaxies and found that all of their bars were consistent with being fast. \citet{font_2017} combined Spitzer images of 68 barred galaxies with previously determined corotation radii to estimate values for $\mathcal{R}$. \citet{cuomo_2019} looked at 16 weakly barred galaxies using data from CALIFA. \citet{guo_2019} used MaNGA data to obtain estimates for pattern speeds, corotation radii and $\mathcal{R}$ for a total of 53 barred galaxies. \citet{garma_oehmichen_2020} combined data from MaNGA and CALIFA to study a sample of 18 galaxies. Finally, \citet{garma_oehmichen_2022} used MaNGA to study 97 barred galaxies.

\begin{figure*}
	\includegraphics[width=\textwidth]{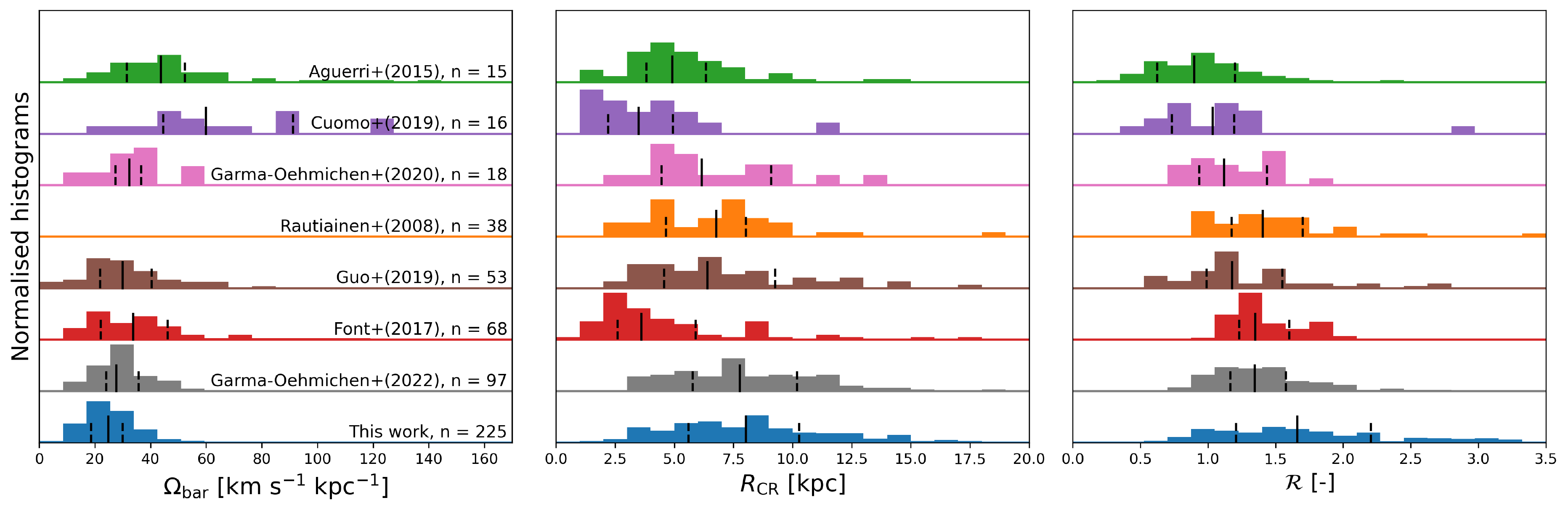}
    \caption{A comparison of estimates of the bar pattern speeds (left), corotation radii (middle) and $\mathcal{R}$ (right) found in various works. All the histograms are normalised and offset from each other vertically to facilitate comparison. The median, 25$^{\rm th}$ and 75$^{\rm th}$ percentile for every distribution are indicated by the short full and dashed lines. The studies are ordered by sample size, which is also shown in the left panel. The values from \citet{rautiainen_2008} and \citet{aguerri_2015} were converted from the observational units cited in their papers to physical units using redshifts obtained from the NASA/IPAC Extragalactic Database (NED). The values obtained from \citet{guo_2019} were converted using their own cited redshifts. Similarly, the values from \citet{font_2017} were converted using the distances cited. The values for the bar pattern speeds from \citet{rautiainen_2008} are not publicly available, hence the empty histogram.}
    \label{fig:comparison}
\end{figure*}

We compare our results with these studies in Figure \ref{fig:comparison}. Our distribution of the bar pattern speed, corotation radius and $\mathcal{R}$ falls well within the range that is usually observed and we see no obvious deviations. Our bar pattern speed distribution agrees especially well with the other studies that have larger sample sizes (n $>$ 50). An interesting trend is observed when looking at the various distributions of $\mathcal{R}$. The median value of $\mathcal{R}$ seems to be moving upwards as the sample sizes increase. This could be attributed to larger samples typically being more representative of large variety of galaxy and bar types, which could have an effect on the observed distribution of $\mathcal{R}$.

Please refer to Appendix \ref{app:comparison_other_work} for a similar comparison to various other studies, but distinguishing between weakly and strongly barred galaxies as well.

\section{Discussion}
\label{sec:discussion}

\subsection{Are strong bars older than weak bars?}

We have found in Figure \ref{fig:omega_hists} that strongly barred galaxies have significantly lower bar pattern speeds than weakly barred galaxies, especially in terms of $\Omega_{\rm b} \sin \left( i \right)$ (p-value $<$ 0.001; $>$3.3$\sigma$). However, there is still a large overlap between the two samples. It is worth noting that this difference is not due to differences in stellar mass, as shown in Figure \ref{fig:against_stellarmass}. Additionally, we found that pattern speed is negatively correlated to bar length in Figure \ref{fig:against_barlen}. This is in agreement with \citet{cuomo_2020}, who used the CALIFA and MaNGA surveys and also found that stronger bars have lower bar pattern speeds. \citet{font_2017} also found that the largest bars have the lowest pattern speeds, while the bars with the largest pattern speeds are all very small. Using CALIFA, MaNGA and Pan-STARRS DR1 (PS1), \citet{lee_2022} also found that the bar pattern speed is negatively correlated to bar length and strength. 

Simulations suggest that the bar grows in size and the pattern speed slows down as the bar exchanges angular momentum with its host \citep{debattista_2000,athanassoula_2003, martinez_valpuesta_2006,okamoto_2015}. Thus, our results suggest that strong bars are older and more evolved structures than weakly barred galaxies. Alternatively, stronger bars could be simply more efficient at redistributing angular momentum, as this depends on various parameters, such as the velocity dispersion and mass distribution of the emitting and absorbing components \citep{athanassoula_2003}.

\subsection{How are strong and weak bars triggered?}

We found that strongly barred galaxies have statistically significantly lower values for $\mathcal{R}$ (p-value = 0.001; 3.3$\sigma$), which was defined as $\mathcal{R} = R_{\rm CR} / R_{\rm bar}$, than weakly barred galaxies, as shown in Figure \ref{fig:r_hist}. However, it is important to note that there is still a big overlap in both distributions. Additionally, we find a relationship between $\mathcal{R}$ and bar length in Figure \ref{fig:against_barlen}, which shows that long bars have lower values for $\mathcal{R}$. However, $\mathcal{R}$ is very dependent on the bar length and inclination estimates \citep{cuomo_2021,roshan_2021a}, so this trend could be primarily caused by differences in bar length between weak and strong bars. A more detailed study using different metrics to measure bar length could help clarify this issue.

We know from simulations that $\mathcal{R}$ depends on the formation of the bar. Bars triggered by tidal interactions tend to have higher values for $\mathcal{R}$ than bars formed by global bar instabilities. Additionally, tidally induced bars stay in the slow regime for a longer time \citep{sellwood_1981,miwa_1998,martinez_valpuesta_2016,martinez_valpuesta_2017}. This seems to suggest that strong bars are more likely to be triggered by bar instabilities, whereas weak bars are more likely to be formed by tidal interactions. This statement could possibly be tested observationally by looking at the environment of a large mass-matched sample of strongly and weakly barred galaxies. Interestingly, this is not in agreement with \citet{cuomo_2020} and \citet{guo_2019}, who found no relationship between bar strength and $\mathcal{R}$, although this could be caused by the lower sample sizes used in these studies.

\subsection{Why and where do we see ultrafast bars?}

Ultrafast bars, which are bars that have $\mathcal{R}$ < 1, should not exist according to our current theoretical understanding. This is because bars are thought to not be able to extend beyond the corotation radius of the galaxy \citep{contopoulos_1980,contopoulos_1981, athanassoula_1992}. Nevertheless, they have been found observationally. Multiple studies find that 26 - 67\% of their bars have $\mathcal{R} < 1$, while 7-40\% have a 1$\sigma$ upper limit that has $\mathcal{R} < 1$ \citep{aguerri_2015, cuomo_2019,guo_2019,garma_oehmichen_2020}. In our sample, $\sim$\todo{11}\% of galaxies have $\mathcal{R} < 1$, while $\sim$\todo{2}\% have a 1$\sigma$ upper limit that has $\mathcal{R} < 1$. This is significantly lower than what others studies typically find. A more detailed breakdown can be found in Table \ref{tab:ultrafast_bars}.

\begin{table*}
    \centering
    \caption{Summary of how many ultrafast bars ($\mathcal{R} < 1$), fast bars ($1 < \mathcal{R} < 1.4$) and slow bars ($\mathcal{R} > 1.4$) are found in various works. Note that \citet{aguerri_2015} and \citet{guo_2019} have multiple different samples, hence the range.}
    \begin{tabular}{lllll}
        \hline
        \hline
         & Sample size & \% Ultrafast & \% Fast & \% Slow \\
        \hline
        Rautiainen et al. (2008) & 38 & 16 & 34 & 50\\
        Aguerri et al. (2015) & 15 & 46-67 & 20-40 & 7-13\\
        Font et al. (2017) & 68 & 1 & 59 & 40\\
        Cuomo et al. (2019) & 16 & 44 & 50 & 6\\
        Guo et al. (2019) & 53 & 26-47 & 13-34 & 38-43\\
        Garma-Oehmichen et al. (2020) & 18 & 39 & 22 & 39\\
        Garma-Oehmichen et al. (2022) & 97 & 11 & 43 & 45 \\
        This work & \todo{225} & \todo{11} & \todo{27} & \todo{62}\\
        \hline
    \end{tabular}
    
    \label{tab:ultrafast_bars}
    \end{table*}

Given that $\mathcal{R}$ is defined as $\mathcal{R} = R_{\rm CR}/R_{\rm bar}$, a low value for $\mathcal{R}$ can arise either because $R_{\rm CR}$ is underestimated or $R_{\rm bar}$ is overestimated. Both seem to happen simultaneously: even though most ultrafast bars have a lower corotation radius (see Figure \ref{fig:Om_Rcr_R}), we see in Figure \ref{fig:R_plot} that ultrafast bars still have a relatively broad range of bar and corotation radii. 

Bars are often associated with spiral arms and rings, so it is often not straight-forward to measure the bar length correctly \citep{hilmi_2020,cuomo_2021,roshan_2021a}. An underestimation of the corotation radius can be a consequence of either an overestimation of the pattern speed, or because the rotation curve is not fitted properly. Additionally, if the inclination of the galaxy is not measured properly, the line of sight velocities in the rotation curve will be corrected incorrectly, which will affect the corotation radius as well. Interestingly, we see slightly more ultrafast bars among strong bars than weak bars (\todo{13.9}\% and \todo{7.8}\%, respectively). 

\subsection{Slow bars}

As noted in Table \ref{tab:ultrafast_bars}, we find that $\sim$\todo{62}\% of galaxies have a slow bar. Our fraction of slow bars is higher than what other studies have typically found. This can be due to multiple factors, such as our larger and more representative sample, which includes weak and strong bars from a volume-limited sample, from a wide range of magnitudes. Additionally, a correct measurement of the bar length is crucial to a correct estimate of $\mathcal{R}$ \citep{cuomo_2021}. Different authors use different methods of measuring bar length, which will change the final distribution of $\mathcal{R}$ (for more details on our bar length measurements, see Section \ref{sec:inputs}). However, perhaps most importantly, \citet{guo_2019} have shown that estimates of the pattern speed will be systematically lower when using kinematic position angles, compared to photometric position angles. This is because the method to calculate the kinematic position angle works by minimising asymmetry in the velocity field, which will reduce the values for the kinematic integrals. This will lower the estimates for the patterns speed, which will, in turn, increase the estimates for $\mathcal{R}$ and produce more slow bars. We used kinematic position angles in this work, which could partially explain the higher fraction of slow bars. Finally, if we take the errors into account, we find that $\sim$\todo{35}\% of our targets have a 1$\sigma$ lower limit that is greater than 1.4. Thus, we can confidently exclude the fast regime for only these targets, which is more consistent with other studies.

\subsection{Strong and weak: part of a continuum}

We have found that strong bars tend to have lower bar pattern speeds. However, as Figure \ref{fig:omega_hists} shows, their distributions overlap significantly. There is no clear threshold in pattern speed, corotation radius or $\mathcal{R}$ that separates weak and strong bars. A closer look at Figure \ref{fig:against_barlen}, where we plot these parameters against bar radius, reveals that the differences in pattern speed and $\mathcal{R}$ are driven by the smallest and largest bars. At intermediate bar radii, the distributions of the two populations overlap. Additionally, the median trend of the weakly and strongly barred subsamples are almost identical in these figures. Figure \ref{fig:rcr_hists} shows that weakly and strongly barred galaxies do not have significantly different corotation radii. As bars are able to grow up until their corotation radius \citep{contopoulos_1980,contopoulos_1981, athanassoula_1992}, this result suggests that weak bars still have the possibility to grow up to the same length as strong bars. Either they have not had to time to do so yet or something else is preventing them.

These results are consistent with the idea of a bar continuum, proposed by \citet{geron_2021}, who found that any distinction in fibre SFR, gas mass and depletion timescale between weak and strong bars disappeared when correcting for bar length. They suggested that weak and strong bars are not fundamentally distinct physical phenomena. Instead, bar types are continuous, and vary from `weakest' to `strongest'. Our measurements of the dynamical parameters of weak and strong bars support this conclusion as well. 

As mentioned in Section \ref{sec:tw_limits}, we require that at least three pseudo-slits can be placed on the bar. This means that this method does not work for the shortest of bars, which are typically weak bars. Thus, it should be kept in mind when interpreting our results that the shortest and weakest bars are not included in our analysis. This manifests itself as well in the errors associated with the measured pattern speeds. The highest errors are associated with bars with the fewest pseudo-slits. This means that weak bars typically have higher errors for the bar pattern speed than strong bars: \todo{6.83} km s$^{-1}$ kpc$^{-1}$ and \todo{4.89} km s$^{-1}$ kpc$^{-1}$, respectively.

\subsection{Dark matter haloes and tension with $\Lambda$CDM}

The fraction of ultrafast and fast bars found in this work is less than what some other studies tend to find, but combined they still comprise \todo{38\%} of our sample. Observing a high fraction of ultrafast and fast bars has been raised as a tension for the $\Lambda$CDM cosmological paradigm, as cosmological simulations predict that bars should slow down significantly \citep{algorry_2017,peschken_2019,fragkoudi_2021,roshan_2021a,roshan_2021b,frankel_2022}. This slowdown of the bar and increase of $\mathcal{R}$ is typically attributed to the dynamical friction applied to the bar by the DM halo \citep{debattista_1998,debattista_2000,fragkoudi_2021}.

As we find fewer ultrafast and fast bars than other studies, this tension is decreased somewhat, but the median value of $\mathcal{R}$ in our sample ($\mathcal{R}$ = $\todo{1.66^{+1.05}_{-0.62}}$) is still significantly lower than that predicted from simulations, whose average values at z$\sim$0 are typically $\mathcal{R}$ > 2.5  \citep{algorry_2017,peschken_2019,roshan_2021b}.  

Other studies have tried to relieve the tension as well. \citet{frankel_2022} has recently shown that simulations obtain higher values of $\mathcal{R}$ than observations, mostly because simulations predict shorter bars, rather than slower bars. Additionally, \citet{fragkoudi_2021} actually do find fast bars in their cosmological simulations in baryon-dominated discs and claim that the DM fraction is too high in other simulations. A lower DM fraction or lower central DM density will lower the dynamical friction, and thus, allow fast bars to exist. Finally, \citet{beane_2022} have shown that the gas phase of the disk can help to stabilise the bar pattern speed and prevent it from slowing down.

As mentioned above, $\mathcal{R}$ is significantly lower for strong bars than for weak bars. This suggests that the DM fractions of strong and weak bars are different as well. Studying the relationship between the DM halo and bars will help us understand the evolution of bars in general. This can be done, for example, with Jeans Anisotropic Modelling (JAM) of these galaxies \citep{cappellari_2008}. This would provide estimates for the DM fraction and allow us to study the intrinsic connection between the DM halo and the dynamical parameters of bars in greater detail.

\subsection{Effect on quenching}

It is also known that strong bars are more often found in red sequence galaxies \citep{masters_2012,vera_2016,cervantessodi_2017,fraser_mckelvie_2020b}. In addition, \citet{geron_2021} showed that stronger bars have the ability to facilitate quenching, whereas weaker bars do not. Figure \ref{fig:deltams_hists} shows that the bar pattern speed and $\mathcal{R}$ are not significantly different between star forming and quenching galaxies. These results suggest that how fast bars rotate, both in terms of pattern speed and $\mathcal{R}$, has no significant or measurable impact on quenching. This seems odd, but one way to look at this is by looking at the timescales involved. A bar with a pattern speed of $\sim$25 km s$^{-1}$ kpc$^{-1}$ will make a full rotation once every $\sim$250 Myr. As (secular) quenching usually happens on $\sim$Gyr timescales \citep{smethurst_2015}, this means that a bar will usually have made multiple full rotations before the galaxy is quenched.

Interestingly, the corotation radius is significantly higher for quenching galaxies. It is known that a bar can grow up to its corotation radius \citep{contopoulos_1980,contopoulos_1981, athanassoula_1992}, so this result relates back to longer and stronger bars having more of an effect on quenching, which agrees with the findings of \citet{geron_2021}.


\section{Conclusion}
\label{sec:conclusion}

We have used the TW method on MaNGA IFU data for a sample of \todo{225} barred galaxies, which is the largest sample this method has been applied to so far. The TW method produces bar pattern speeds, which is used to calculate corotation radii and $\mathcal{R}$, the ratio between the corotation radius and bar radius. We have used Galaxy Zoo morphological classifications to distinguish between strongly and weakly barred galaxies. This allows us to study the bar pattern speed, corotation radius and $\mathcal{R}$ for a statistically significant sample of strongly and weakly barred galaxies, and compare them with each other. We have found that:

\begin{itemize}
    \item Though there is significant overlap, we find that the bar pattern speeds between weakly and strongly barred galaxies are significantly different, as the p-value of the comparisons in physical units is $<$0.001-0.002 (which corresponds to 3.1-3.3$\sigma$). The median bar pattern speed of strongly barred galaxies ($\Omega_{\rm b} = \todo{23.36^{+9.25}_{-8.1}}$ km s$^{-1}$ kpc$^{-1}$) is lower than that of weakly barred galaxies ($\Omega_{\rm b} = \todo{25.91^{+10.42}_{-7.26}}$ km s$^{-1}$ kpc$^{-1}$). Additionally, we find that the bar pattern speed is inversely proportional to bar length. We also show that this difference is not due to differences in stellar mass in our targets. Simulations suggest that the pattern speed goes down as the bar evolves and exchanges angular momentum, so our results suggest that strong bars are older and more evolved structures than weak bars.

    \item We could not find evidence that the corotation radius between weakly (R$_{\rm CR} = \todo{7.19^{+3.82}_{-2.96}}$ kpc) and strongly barred galaxies (R$_{\rm CR} = \todo{8.33^{+4.57}_{-3.31}}$ kpc) is significantly different, as the p-value of the comparison in physical units is 0.012 (which corresponds to 2.5$\sigma$). As bars can grow up until their corotation radius, this result suggests that weak bars still have the possibility of becoming as long as strong bars.
    
    \item Despite the significant overlap in the distributions, we find that $\mathcal{R}$ is statistically significantly lower for strong bars than for weak bars (p-value = 0.001; 3.3$\sigma$). The median value for strong bars is $\todo{1.53^{+0.74}_{-0.53}}$, while it is $\todo{1.88^{+1.08}_{-0.75}}$ for weak bars. Additionally, we find that $\mathcal{R}$ is inversely proportional to bar length and that these differences are not caused by differences in stellar mass. As $\mathcal{R}$ is related to the formation of bars, this suggests that weak bars are more likely to be formed by tidal interactions, whereas strong bars are more likely to be triggered by global bar instabilities.
    
    \item We do not see a distinct cutoff or threshold in pattern speed, corotation radius or $\mathcal{R}$ that separates strong and weak bars from each other. In fact, the overlap is still quite significant. This is consistent with \citet{geron_2021}, who stated that strong and weak bars are not distinct physical phenomena, but rather lie on a continuum of bar types that vary from `weakest' to `strongest'.
    
    \item \todo{11}\% of our sample host ultrafast bars, \todo{27}\% host fast bars and \todo{62}\% of all bars are slow. We have a slightly higher fraction of slow bars and a lower fraction of ultrafast bars than most other studies. This can be attributed to various factors, such as the bigger and more representative sample used in this study.
    
    \item However, only $\sim$\todo{2}\% of our targets have a 1$\sigma$ upper limit that has $\mathcal{R} < 1$. Similarly, we can only confidently exclude the (ultra)fast regime for $\sim$\todo{35}\% of our galaxies (i.e. they have a 1$\sigma$ lower limit that has $\mathcal{R} > 1.4$)
    
    \item The lower fraction of ultrafast bars among our sample decreases the recent tension with $\Lambda$CDM. However, the median value of $\mathcal{R}$ in our sample is still significantly lower than what is predicted from simulations.
    
    \item We do not see any significant difference between the star forming and quenching subsamples in terms of pattern speed or $\mathcal{R}$. However, quenching galaxies do have significantly higher corotation radii than star forming galaxies.

\end{itemize}

\section*{Acknowledgements}
The data in this paper are the result of the efforts of the Galaxy Zoo volunteers, without whom none of this work would be possible. Their efforts are individually acknowledged at \url{http://authors.galaxyzoo.org}.

The authors of this paper are grateful for the helpful comments of M. Bureau and Y. Zou.

Funding for the Sloan Digital Sky Survey IV has been provided by the Alfred P. Sloan Foundation, the U.S. Department of Energy Office of Science, and the Participating Institutions. SDSS acknowledges support and resources from the Center for High-Performance Computing at the University of Utah. The SDSS web site is www.sdss.org.

Funding for the Sloan Digital Sky Survey IV has been provided by the Alfred P. Sloan Foundation, the U.S. Department of Energy Office of Science, and the Participating Institutions. 

SDSS-IV acknowledges support and resources from the Center for High Performance Computing  at the University of Utah. The SDSS website is www.sdss.org.

SDSS-IV is managed by the Astrophysical Research Consortium for the Participating Institutions of the SDSS Collaboration including the Brazilian Participation Group, the Carnegie Institution for Science, Carnegie Mellon University, Center for Astrophysics | Harvard \& Smithsonian, the Chilean Participation Group, the French Participation Group, Instituto de Astrof\'isica de Canarias, The Johns Hopkins University, Kavli Institute for the Physics and Mathematics of the Universe (IPMU) / University of Tokyo, the Korean Participation Group, Lawrence Berkeley National Laboratory, Leibniz Institut f\"ur Astrophysik Potsdam (AIP),  Max-Planck-Institut f\"ur Astronomie (MPIA Heidelberg), Max-Planck-Institut f\"ur Astrophysik (MPA Garching), Max-Planck-Institut f\"ur Extraterrestrische Physik (MPE), National Astronomical Observatories of China, New Mexico State University, New York University, University of Notre Dame, Observat\'ario Nacional / MCTI, The Ohio State University, Pennsylvania State University, Shanghai Astronomical Observatory, United Kingdom Participation Group, Universidad Nacional Aut\'onoma de M\'exico, University of Arizona, University of Colorado Boulder, University of Oxford, University of Portsmouth, University of Utah, University of Virginia, University of Washington, University of Wisconsin, Vanderbilt University, and Yale University.

TG gratefully acknowledges funding from the University of Oxford Department of Physics and the Saven Scholarship.

RJS gratefully acknowledges funding from Christ Church, University of Oxford.

BDS acknowledges support through a UK Research and Innovation Future Leaders Fellowship [grant number MR/T044136/1].

This project makes use of the MaNGA-Pipe3D dataproducts. We thank the IA-UNAM MaNGA team for creating this catalogue, and the Conacyt Project CB-285080 for supporting them. 

\section*{Data availability}

The Galaxy Zoo data used in this article will be made publicly available soon. The data from MaNGA is publicly available and can be found at http://www.sdss.org/dr17/manga/. The Legacy survey can be accessed at https://www.legacysurvey.org/. The code used in this paper as well as the data in Tables 1 and 3 can be found at https://doi.org/10.5281/zenodo.7567945.





\bibliographystyle{mnras}
\bibliography{bibtex.bib}



\appendix

\section{Comparison to other work}
\label{app:comparison_other_work}

\begin{figure*}
	\includegraphics[width=\textwidth]{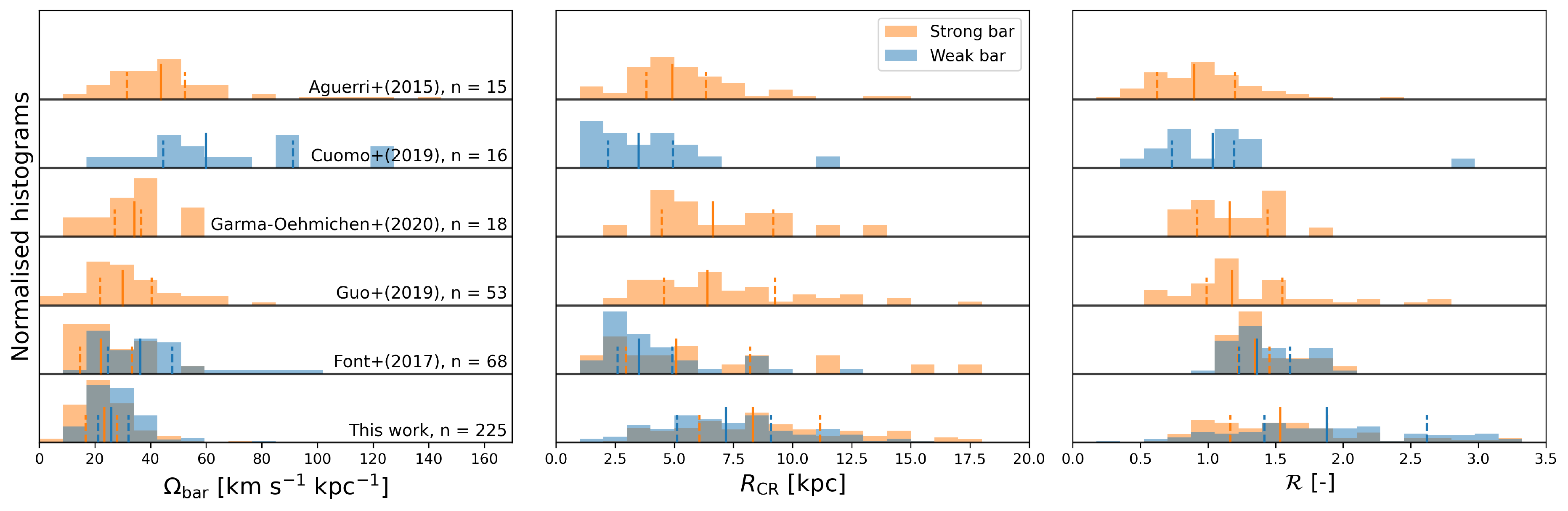}
    \caption{A comparison of estimates of the bar pattern speeds (left), corotation radii (middle) and $\mathcal{R}$ (right) found in various works. The data is split between weakly barred galaxies (blue) and strongly barred galaxies (orange). \citet{font_2017} have both weak and strong bars, whereas the other studies focus only on either weak or strong bars. All the histograms are normalised and offset from each other vertically to facilitate comparison. The median, 25$^{\rm th}$ and 75$^{\rm th}$ percentile for every distribution are indicated by the short full and dashed lines. The studies are ordered by sample size, which is also shown in the left panel. The values from \citet{aguerri_2015} were converted from the observational units cited in the papers to physical units using redshifts obtained from the NASA/IPAC Extragalactic Database (NED). The values obtained from \citet{guo_2019} were converted using their own cited redshifts. Similarly, the values from \citet{font_2017} were converted using the distances cited in the paper.}
    \label{fig:comparison_WBSB}
\end{figure*}

In Section \ref{sec:comparison_other_work}, we compared our results to that of other studies. However, we did not make a distinction between weak and strong bars. In Figure \ref{fig:comparison_WBSB}, we compare our results to that of various other works that clearly mentioned whether the bars studied were weak or strong. We see that, in all the other studies, the bar pattern speed for weakly barred galaxies tends to be higher than that of strongly barred galaxies, which corresponds to our findings. Contrary to our findings, strongly barred galaxies in the other studies do seem to have higher corotation radii than weakly barred galaxies. However, as mentioned in Section \ref{sec:calc_rcr_r}, \citet{aguerri_2015}, \citet{cuomo_2019} and \citet{guo_2019} calculate the corotation radius assuming that it lies in the region where the rotation curve has flattened, which is not always the case and will bias the final values. Finally, weakly and strongly barred galaxies seem to have comparable values for $\mathcal{R}$ in the other studies, which is in contrast to what we have found.

\bsp	
\label{lastpage}
\end{document}